# Radiation Damage in Silicon Detectors
## Caused by Hadronic and Electromagnetic Irradiation

E. Fretwurst[1], G. Lindstroem[1], I. Pintilie[1,2], J. Stahl[1]

[1]University of Hamburg, Germany,
[2]National Institute for Material Physics, Bucharest

# Contents

# Introductory Note

Segmented silicon detectors (microstrip and pixel devices) are widely used in High Energy Physics applications. Due to their unsurpassed quality they play a major role in the tracking area of present and future collider experiments. However in the forthcoming experiments at LHC and TESLA they are subject to extremely harsh hadronic resp. electromagnetic radiation environments thus threatening their operability.

In the past years considerable effort had been undertaken in order to improve the radiation tolerance of these devices and thus guarantee an operational lifetime of e.g. 10 years in the LHC inner detector. Most of this work had been carried out by the CERN-RD48 (ROSE) collaboration (1996-2000). It had been shown that radiation damage effects in silicon detectors could be considerably reduced by using a defect engineering approach, consisting in an oxygen enrichment of the float zone wafers during detector processing. This technique is now employed e.g. in the ATLAS pixel and part of the strip layers. In future experiments like for an upgraded LHC (SLHC) the present luminosity of $10^{34}$ is envisioned to be increased by a factor of 10 and hence a hadronic fluence integrated over the operational period will reach $10^{16}$ equivalent n/cm². Presently there are no techniques available to ensure the necessary radiation tolerance of tracking detectors up to such high values. Possible improvements of the ROSE results as well as completely new approaches are necessary and will be the goal of the newly founded R&D collaboration CERN-RD50.

Our group is engaged in long term R&D projects in the development and radiation hardness of silicon detectors. The group participated in RD48 and is presently also a member of RD50. Our main object is the systematic investigation of damage consequences for the operation of silicon detectors and its complete understanding. Although considerable improvement had been achieved in the systematic analysis of radiation induced changes of macroscopic detector properties only little was gained so far in the correlation with responsible defects on the microscopic scale. The first paper in this report gives an overview of the present knowledge as resulting largely from RD48-experiments and recent own upgrades for hadron induced damage. Both confirmed results showing the improvements to be gained by the oxygen enrichment as also still open questions in the available systematics are covered. As hadronic interactions give rise to both "cluster" and "point defects" in silicon, low energy electromagnetic irradiation can only produce point defects, offering a much more direct correlation between defect generation and detector performance. All other 3 papers are dealing with this issue. In the second one we investigate existing point defects in as processed detectors and additional defect formation after gamma irradiation. The last two paper are concentrating on a close correlation between gamma induced deep defects and their consequences for the detector performance. For the very first time it is shown that both the change of the effective doping concentration (depletion voltage) and the free charge carrier generation (reverse current) can be completely understood by the microscopically investigated formation of deep defects, largely suppressed by oxygen enrichment. This is regarded to be a major breakthrough for the long searched for understanding of radiation damage and will hopefully also pave the road for an optimum of radiation tolerant detectors including hadronic damage.

The Work of our group is supported by the DFG (German Research Foundation) and BMBF (Ministry for Research). Part of the work had been performed in collaboration with industry (CiS Erfurt), supported by the BMWI (Ministry for Economy).

# Radiation Damage in Silicon Detectors[†]

## Gunnar Lindström[*]

*Institute for Experimental Physics, University of Hamburg, 22761 Hamburg, Germany*

**Abstract**

Radiation damage effects in silicon detectors under severe hadron- and $\gamma$–irradiation are surveyed, focusing on bulk effects. Both macroscopic detector properties (reverse current, depletion voltage and charge collection) as also the underlying microscopic defect generation are covered. Basic results are taken from the work done in the CERN-RD48 (ROSE) collaboration updated by results of recent work. Preliminary studies on the use of dimerized float zone and Czochralski silicon as detector material show possible benefits. An essential progress in the understanding of the radiation induced detector deterioration had recently been achieved in gamma irradiation, directly correlating defect analysis data with the macroscopic detector performance.

# 1. Introduction

Silicon pixel and microstrip detectors are today's best choices for meeting the requirements of tracking applications in the forthcoming LHC experiments. However, due to the extreme harsh radiation environment the required prolonged operability poses an unprecedented challenge to their radiation tolerance. At the expected yearly level of up to several $10^{14}$ hadrons per cm² during the foreseen 10 years of operation the damage induced changes in the silicon bulk will cause an increase of both the reverse current and the necessary depletion voltage as well as a decrease of the charge collection efficiency. Using standard silicon material and present detector technology these damage induced effects lead to not tolerable deteriorations in the detector performance. Possible strategies for improving the radiation hardness of these devices consist of using suitable defect engineering of the silicon material, as introduced by the CERN-RD48 (ROSE) collaboration [1], operation at very low temperatures under investigation by the CERN-RD39 group [2] or a specialized design of detector geometry, presently proposed or under test by several other groups, see e.g. [3] and literature cited there.

This report is based on the results of the ROSE collaboration showing the appreciable improvements which had been achieved by diffusing oxygen from the SiO₂-Si interface into the silicon bulk after initial oxidation of the silicon wafers. This so called DOFZ (Diffusion Oxygenated Float Zone) technology had proven to be the most cost effective way of adequately modifying the standard process technology (see section 4.1).

In the following different aspects of hadron and gamma/lepton induced bulk damage effects, originating from the Non Ionizing Energy Loss (NIEL) are summarized, especially comparing standard with oxygenated silicon detectors, as extracted from numerous publications of the ROSE collaboration [4]. In contrast to bulk effects surface and interface related changes, produced by the ionizing energy loss (radiation dose), play a minor role for lifetime limitations of silicon detectors. The discussion of such effects can e.g. be found in [5].

# 2. Primary damage effects

The bulk damage in silicon detectors caused by hadrons or higher energetic leptons respectively gammas is primarily due to displacing a Primary Knock on Atom PKA out of its lattice site. The threshold energy for this process is ~ 25 eV. Such single displacements resulting in a pair of a silicon interstitial and a vacancy (Frenkel pair) can be generated by e.g. neutrons or electrons with an energy above 175 eV and 260 keV respectively. Low energy recoils above these threshold energies will usually create fixed point defects. However for recoil energies above about 5 keV a dense agglomeration of defects is formed at the end of the primary PKA track. Such disordered regions are referred to as defect clusters. The kinematic lower limits for the production of clusters are ~35 keV for neutrons and ~8 MeV for electrons. In figure 1 the simulation result for a 50 keV PKA (average recoil produced by 1 MeV neutrons) is shown, clearly distinguishing between discretely distributed vacancies (point defects) and dense agglomerates (clusters) [6]. It should be noted here that the radiation damage caused by Co-60 gammas is primarily due to the interaction of Compton electrons having a maximum energy of only 1 MeV. Hence in this case cluster production is not possible and the damage is exclusively due to point defects.

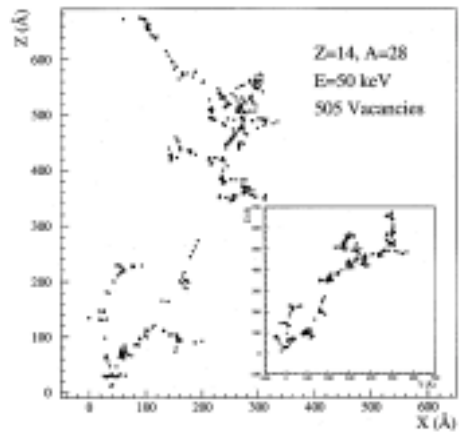

Fig. 1. Simulation results: Sample distribution of vacancies induced by a 50 keV recoil silicon atom; insert shows transverse projection [6].

Both point defects and clusters can have severe effects on the detector performance, depending on their concentration, energy level and the respective electron and hole capture cross section. Defects with deep energy levels in the middle of the forbidden gap could act as recombination/generation centers and are hence responsible for an increase of the reverse detector current. The removal of dopants by formation of complex defects as well as the generation of charged centers changes the effective doping concentration and the needed operating voltage to fully deplete the detector thickness. Finally such defects could also act as trapping centers affecting the charge collection efficiency. In view of the LHC operational scenario it has also to be noted that the primary produced defects may be subject to changes after long term storage at room temperature. Such annealing effects may very likely be caused by changes due to the dissolution of clusters releasing migrating vacancies and interstitials [7]. The observed time constants for such "reverse annealing" processes are rather large (for standard silicon at room temperature e.g. ~ 1 year). A reliable projection of damage results on the LHC operational scenario is hence only possible if annealing effects are included comparable to those expected in 10 years of LHC operation. This task had been successfully attacked by performing studies at elevated temperatures thus accelerating the kinetic effects being responsible for such long term changes. Related basic studies are found in [8-10].

## 3. NIEL scaling of bulk damage

The preceding discussion and numerous experimental observations have led to the assumption that damage effects produced in the silicon bulk by energetic particles can be described as being proportional to the so called displacement damage cross section D. This quantity is equivalent to the <u>N</u>on <u>I</u>onizing <u>E</u>nergy <u>L</u>oss (NIEL) and hence the proportionality between the NIEL-value and the resulting damage effects is referred to as the NIEL-scaling hypothesis (for deviations to this rule see section 4.5.1). The displacement damage cross section D is normally quantified in MeVmb, whereas the NIEL-value is given in keVcm²/g. For silicon

with A=28.086 g/mol the relation between D and NIEL is:

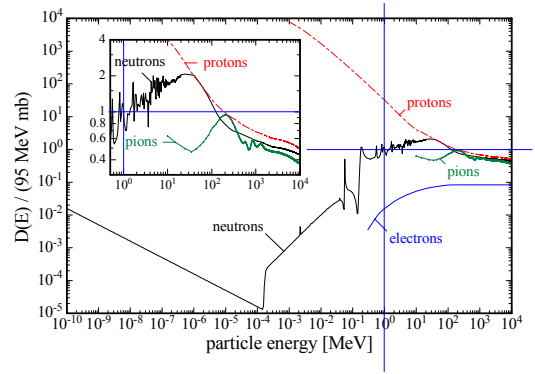

Fig. 2. Non ionising energy loss NIEL for different particles.

100 MeVmb = 2.144 keVcm²/g. The D or NIEL value is depending on the particle type and energy. According to an ASTM standard, the displacement damage cross section for 1 MeV neutrons is set as a normalizing value: $D_n(1\text{MeV}) = 95$ MeVmb [11]. On the basis of the NIEL scaling the damage efficiency of any particle with a given kinetic energy can then be described by the hardness factor κ. Applying the NIEL hypothesis, one may replace the actual particle energy spectrum dΦ/dE by a NIEL folded spectrum and the damage effect, caused by its total fluence $\Phi_p$ by a 1MeV neutron equivalent fluence $\Phi_{eq} = \kappa \cdot \Phi_p$, a more detailed discussion is found in [12, 13].

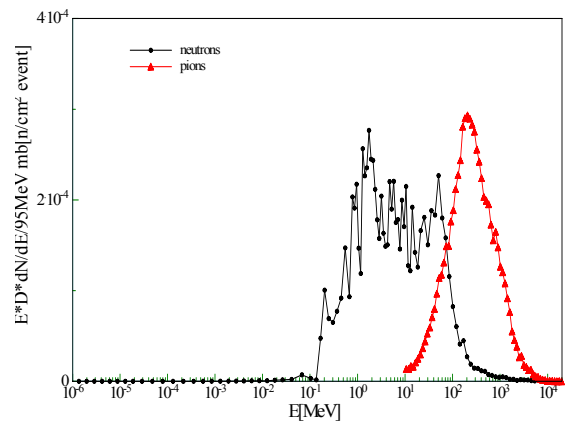

Fig. 3. NIEL folded energy spectra for neutrons and pions in the SCT region (silicon counter tracker) of the LHC ATLAS detector.

In figure 2 the normalized NIEL values are plotted as function of energy. The data are taken from [14-17]. More recent calculations can be found in [6], [18]. NIEL scaling and its limitations is extensively discussed in [6]. Regardless of possible deviations in certain cases, the NIEL scaling should always be applied as a first order approximation of the damage efficiency. Hence it is useful to display the NIEL folded energy spectra for the most abundant particles (neutrons and pions) in the LHC experiments, as done for the SCT region of ATLAS in figure 3 [12]. Finally the resulting equivalent fluences as function of radius are shown in figure 4.

From figure 3 we conclude that reactor neutrons, ranging mainly from 1 to 10 MeV, are adequate for reliable damage tests and that indeed irradiations with 250 MeV pions, available at PSI-Villigen, should result in similar damage as expected in LHC. Many of the past and present irradiation tests have however been performed using the 23 GeV proton beam at CERN-PS. In view of reliable predictions for the LHC performance it is therefore reassuring that damage results from 23 GeV proton and 250 MeV pion irradiation agree pretty well, if proper NIEL scaling is used.

## 4. Survey of damage results

### 4.1. DOFZ technology

The key idea of the RD48 strategy implied that the radiation tolerance of silicon can be improved by adequate defect engineering. Defect engineering involves the deliberate addition of impurities in order to reduce the damage induced formation of electrically active defects. Based on early attempts at the Instrumentation Division of BNL [19], RD48 relied on carbon and oxygen as the key ingredients for reaching this goal [1]. A wide range of oxygen and carbon concentrations had been investigated using several different process technologies.

It was finally found that an oxygen enrichment can best be achieved by a post-oxidation diffusion of oxygen into the silicon bulk. This DOFZ process (<u>D</u>iffusion <u>O</u>xygenated <u>F</u>loat <u>Z</u>one) has been performed at temperatures between 1100 and 1200°C and with durations between several hours and 10 days, reaching an oxygen concentration between 1 and $4 \cdot 10^{17}$ O/cm³ (see also figure 23).

Practical applications are employing a diffusion at 1150°C for 24-72h. The O-concentration reached this way is around $2 \cdot 10^{17}$ O/cm³, see figure 5.

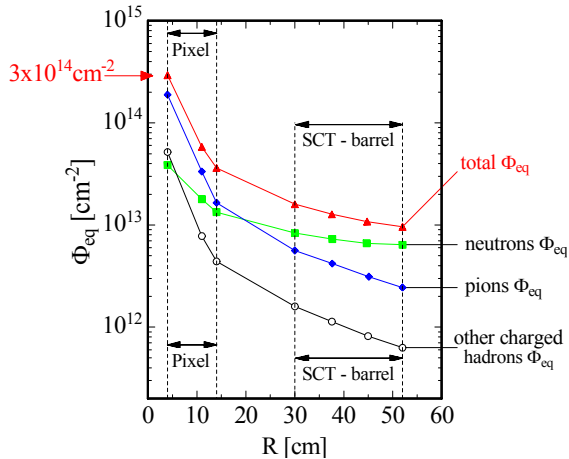

Fig. 4. Hadron fluences expected in the ATLAS inner detector.

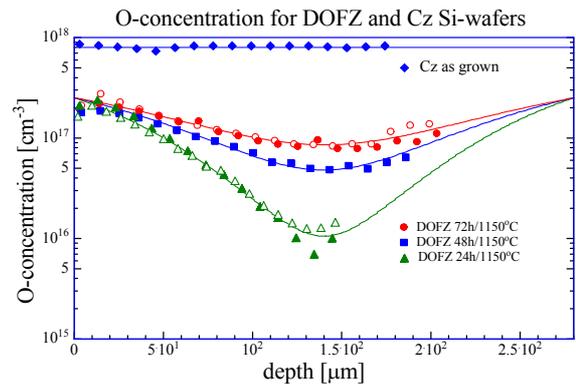

Fig. 5. Depth profiles of O-concentration for three different DOFZ processes, from bottom: 24, 48, 72 h at 1150 °C; in comparison: Czochralski silicon; measured by SIMS [20, 21].

It should be emphasized that the DOFZ process can easily be included in the standard detector process technology and had proven to be a most cost effective technique. It was meanwhile successfully transferred to several manufacturing companies, see [22] and is e.g. employed in the design and processing of pixel and microstrip detectors for the ATLAS experiment [23].

### 4.2. bulk generation current

The damage induced increase of the bulk generation current $\Delta I$ exhibits a quite simple dependence on the irradiating particle type and fluence.

If normalized to the sensitive volume V (best defined by using a guard ring structure around the diode pad), the current increase is strictly proportional to the 1 MeV neutron equivalent fluence $\Phi_{eq}$ (see section 3), as demonstrated in figure 6 [10]. Therefore a "current related damage rate $\alpha$" can be defined by:

$$\Delta I/V = \alpha \cdot \Phi_{eq} \qquad (1)$$

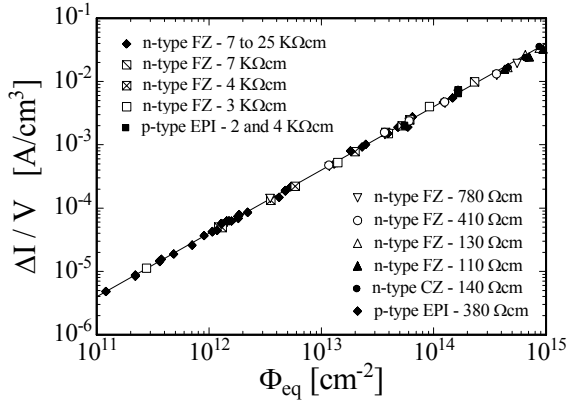

Fig. 6. Damage induced bulk current as function of particle fluence for different detector types.

The measured value of the reverse current depends exponentially on the operating temperature in a straight forward way and $\alpha$-values are therefore always normalized to 20°C. Details for this normalization and the temperature dependent beneficial annealing, depicted in figure 7, can be found in [10, 24]. At any given temperature and annealing time the damage rate $\alpha$, if temperature normalized, is a universal constant, not depending on the material type (n- or p-type, FZ, epi or Cz silicon, resistivity), or irradiating particles (neutrons, protons, pions). Therefore $\alpha$ is often used to reliably monitor the accumulated particle fluence. A recommendation for this task is its measurement after an annealing of 80 min at 60°C. In this situation $\alpha$ is independent of the detailed history of the actual irradiation. The ROSE adapted value is: $\alpha_{80/60} = 4.0 \cdot 10^{-17}$ A/cm. For an annealing temperature of 80°C the same value would be reached after an annealing time of ~10 minutes [10, 25].

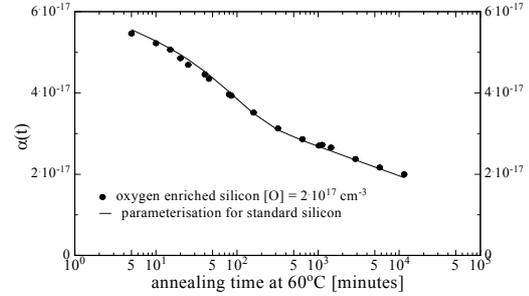

Fig. 7. Annealing function of current related damage parameter $\alpha$ for standard (solid line) and DOFZ (dots) silicon detectors.

It should be emphasized that the bulk current increase and its annealing function follow strictly the NIEL scaling. This and the observation, that the large current values cannot be explained by any known discrete point defects led to the assumption of a cluster related effect. The correlation with the enhanced density of double vacancies in clusters as observed in [25] and a possible interstate charge transfer mechanism, as proposed in [26], are likely explanations.

### 4.3. effective doping and depletion voltage

The depletion voltage $V_{dep}$, necessary to fully extend the electric field throughout the depth d of the detector (asymmetric junction) is related with the effective doping concentration $N_{eff}$ of the silicon bulk

by

$$V_{dep} = (q_0/2\varepsilon\varepsilon_0)|N_{eff}|d^2 \qquad (2)$$

This equation holds not only for the original n-type silicon (donor doped) but also after severe irradiation when the effective doping concentration changes its sign by increased generation of "acceptor like" defects. In any case one can describe the change of the depletion voltage by the related change in $|N_{eff}|$ = $|N_d - N_a|$ with $N_d$ and $N_a$ as the positively charged donor and negatively charged acceptor concentration.

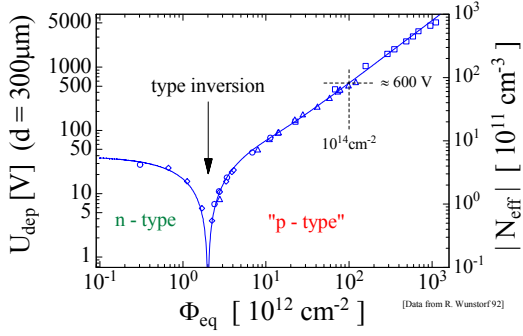

Fig. 8. Change of the effective doping concentration in standard silicon, as measured immediately after neutron irradiation.

The result of a first systematic study of the extremely large change of $|N_{eff}|$ as measured immediately after irradiation of standard silicon detectors is depicted in figure 8 [8]. The not tolerable increase of the needed depletion voltage after "type inversion" had then been the major impact for radiation hardening.

### 4.3.1. CERN-scenario measurements

In real applicational scenarios damage induced "immediate" changes of $N_{eff}$ as displayed in figure 8 will always be connected with subsequent "annealing" effects. It was therefore found useful to devise an almost online possibility for rapidly comparing the radiation tolerance between different materials, taking the yearly expected annealing into account. The so called CERN-scenario experiments consist of consecutive irradiation steps with a short term annealing in between [27]. The annealing steps (4 minutes at 80°C) are chosen such that the observed change in $N_{eff}$ corresponds to what would be expected after about 10 days at 20°C. Thus the operational LHC scenario can be approximated this way. The chosen annealing time corresponds also to the minimum of the annealing curve as measured for 80°C and is therefore closely related to the "stable component" of the change in $N_{eff}$ (see below).

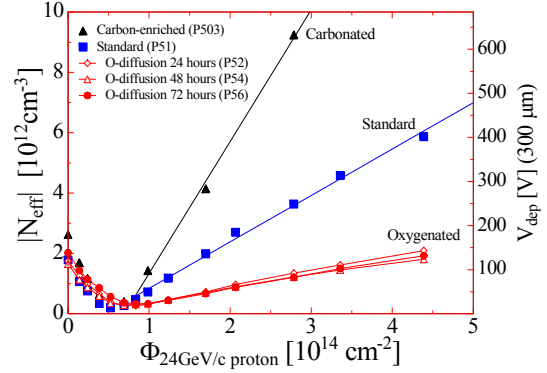

Fig. 9. Influence of carbon and oxygen enrichment to changes of the effective doping concentration after proton irradiation.

Figure 9 shows a comparison between standard, carbon- and oxygen enriched silicon as irradiated by 23 GeV protons at the CERN PS irradiation facility [27]. It is clearly seen, that the O-enrichment reduces the change in Neff and likewise in the depletion voltage considerably to about 1/3 of that for standard silicon while carbon-enrichment proves to have an adverse effect. However, although pion and proton irradiation exhibit the same superiority of DOFZ material, neutron damage does not result in a similar improvement, see figure 10 [27]. This discrepancy with the simple NIEL scaling assumption (section 3) is often been referred to as the "p-n-puzzle". A discussion of this point will be given in section 4.5.1.

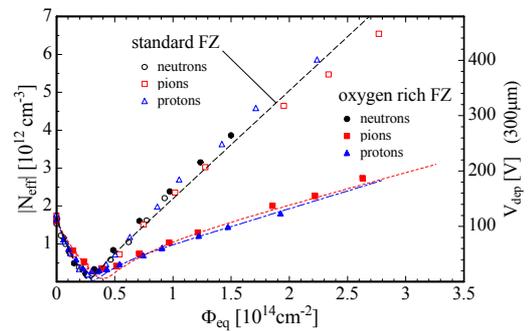

Fig. 10. Particle dependance of radiation damage for standard and oxygenated silicon detectors.

### 4.3.2. Annealing experiments and modeling

As already mentioned, the CERN scenario tests allow only the measurement of one relevant damage parameter, namely the change of the effective doping concentration around the minimum of the annealing function. A different but much more time consuming approach, introduced by the Hamburg group, uses a set of diodes which are irradiated individually with different fluences and are then subjected to full annealing cycles at an elevated temperature for accelerating the annealing kinetics (see [13] and literature cited there). The result of such an annealing experiment is shown in figure 11. $\Delta N_{eff}$ is the damage induced change in the effective doping concentration:

$$\Delta N_{eff}(\Phi_{eq}, t(T_a)) = N_{eff,0} - N_{eff}(\Phi_{eq}, t(T_a)) \qquad (3)$$

As function of the equivalent fluence $\Phi_{eq}$ and the annealing time t at temperature T $\Delta N_{eff}$ can be described as:

$$\Delta N_{eff} = N_A(\Phi_{eq}, t(T_a)) + N_C(\Phi_{eq}) + N_Y(\Phi_{eq}, t(T_a)) \qquad (4)$$

$\Delta N_{eff}$ consists of three components, a short term beneficial annealing $N_A$, a stable damage part $N_C$ and the reverse annealing component $N_Y$. Both time constants for the beneficial as well as reverse annealing depend strongly on the temperature. A more detailed discussion can be found in [10, 13]. Recent results for the stable damage component $N_C$ have shown that there is no pronounced difference between a moderate 24 h and an enlarged 72 h oxygen diffusion and in fact the improvement with respect to standard material is also only marginal as to be seen in figure 12 [20]. This surprising effect will be further discussed in section 4.6.

A much more general behavior regarding the DOFZ benefits is seen for the reverse annealing amplitude. Standard silicon reveals an almost linear behavior of the reverse annealing amplitude $N_Y$ with fluence while DOFZ silicon shows strong saturation at large fluences. This effect as well as an observed increased time constant [1, 20] offer an additional safety margin for detectors kept at room temperatures every year during maintenance periods. As shown in [20] and documented in figure 13, the difference between 24 and 72 h O-diffusion is again not very large. It is however interesting that there are, contrary to previous investigations, also some benefits after neutron irradiation.

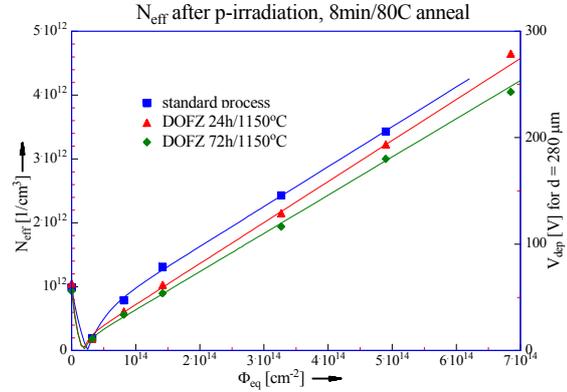

Fig. 12. Stable damage related $N_{eff}$ in minimum of annealing curve after PS-p-irradiation for standard and DOFZ silicon.

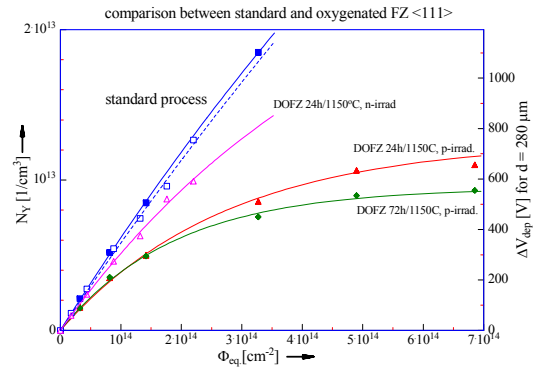

Fig. 13. Reverse annealing amplitude as function of equivalent fluence after n- (open symbols) and PS-p-irradiation (filled symbols). For neutrons the data are normalized to those for protons for the standard process [20].

With all parameters of eq. (4) it is then possible to calculate the expected change of the necessary depletion voltage throughout prolonged operation in any given hadron radiation environment. Such projections on the 10 year LHC operational scenario are finally shown in figure 14. It is obvious that only the DOFZ method guarantees the required lifetime expectancy of 10 years for the innermost pixel layer of the ATLAS experiment (see also [23]). It should however be emphasized that the damage projection shown here is based on the parameters listed in [1] thus may underestimate the beneficial DOFZ effect according to recent results as shown in figure 13 (see also [20]). The actual values seem also to depend on the individual manufacturing process (see section 4.5). It should also be mentioned that annealing experiments have to include a long known bi-stable effect [28]. Thus final conclusions can only be drawn after some proper quality assurance check of the actual detectors.

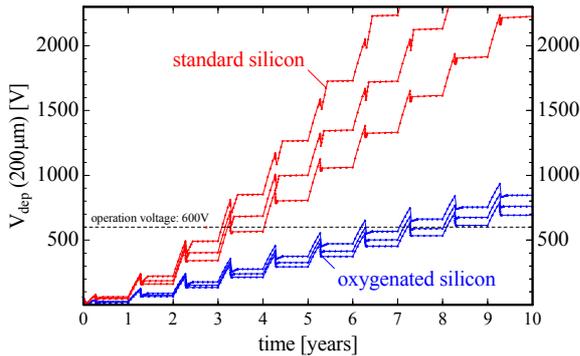

Fig. 14. Damage projections for the ATLAS pixel detector at r = 4cm (B-layer) as an example. Curves from bottom to top: yearly maintenance period at 20ºC: 14, 30 and 60 days [1].

### 4.4. Charge collection efficiency

For any application of silicon detectors in vertex detectors the prime focus is on the operating voltage needed to guarantee a sufficiently good S/N ratio for detection of mip's. The signal noise is controlled by operating the detectors at low temperature (–10°C). Such, a good charge collection efficiency is then the prime objective. For best performance full depletion of the detector thickness is needed and therefore the considerations outlined in section 4.3 are of great importance. However, in order to keep trapping and ballistic deficit effects low, the operating voltage has to be even larger than the depletion voltage.

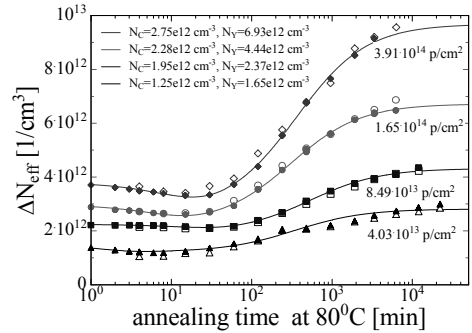

Fig. 15. Annealing curves as measured from charge collection (open symbols) and C/V-measurements (closed ones).

Measurements of charge collection in test diodes have been performed with β-sources and long wavelength infrared laser light pulses, which ensure a near mip like charge track generation throughout the depth of the device (see e.g. [29]). Other experiments have used short wavelength IR laser beams illuminating the test diode from either $p^+$ or $n^+$ side for investigation of the electron resp. hole dominated pulse signals separately. This allows the study of charge carrier trapping effects and these results could then again be included in model calculations for LHC or other applications [30-33].

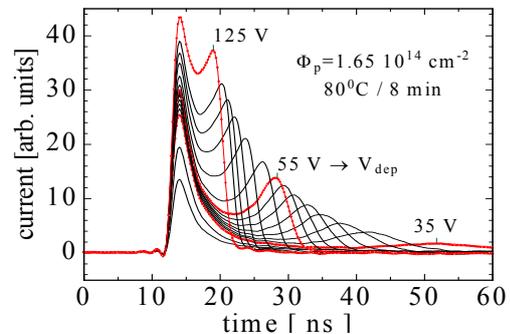

Fig. 16. Time resolved diode signals for 1 nsec IR-laser injection (λ=830 nm) on $p^+$-electrode, after 1.65·10¹⁴ p/cm² irradiation [1].

The analysis of charge collection as function of bias voltage should, similar to capacitance/voltage measurements, also reveal the depletion voltage. It is reassuring that in fact both techniques have led to the same results (figure 15) [1]. Examples of time resolved signal pulse shapes are shown in figure 16 [1]. The double peaking is attributed to a double junction structure of the kind p$^+$npn$^+$. This is a generally observed effect and still needs its complete explanation, a likely explanation as also other details of the signal shape are discussed in [34, 35].

The application of charge collection results as obtained with test diodes to segmented detectors is not straight forward but incorporates a more complex calculation involving the geometrical structure of the detector by using a "weighting field". This point had long been neglected and only recently caught more attention, for detailed discussion see [36].

### 4.5. Microscopic understanding of damage effects

Systematic studies for defect engineering are finally only possible if the correlation between defect formation and the detector performance would be known. Numerous investigations have been undertaken for characterization of damage induced defects in detector grade silicon, especially using DLTS (Deep Level Transient Spectroscopy) and TSC (Thermally Stimulated Current) techniques, showing a lot of relevant results [9, 10, 25, 37-40]. A promising correlation has been reported between the cluster formation as visible in the double vacancy defect and the damage related current [25]. Other studies have given a lot of detailed information of defect kinetics visible in the changes during prolonged annealing or as function of temperature (see e.g. [9, 10, 37]). However despite many partially useful attempts it was so far not possible to quantitatively fully understand the (macroscopic) detector performance on the basis of such (microscopic) data. In the following two examples are given, which are regarded to be promising in this respect.

### 4.5.1. p-n puzzle, violation of NIEL scaling

As outlined in section 3, originally all damage effects had been assumed to be simply proportional to the displacement damage cross section and hence to obey a universal behavior not depending on the particle type and energy if NIEL scaling is applied. NIEL scaling is still valid for the damage induced bulk generation current of high energy hadron irradiated detectors (section 4.2) but is strongly violated for the change of the effective doping concentration (see section 3, figure 10 and 13).

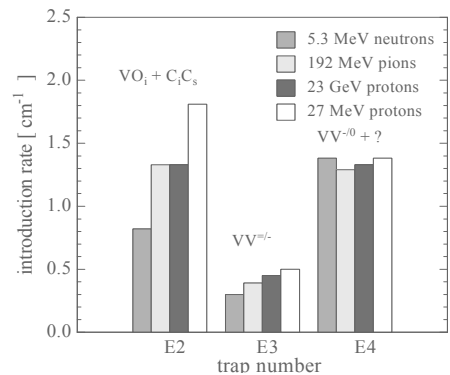

Fig. 17. Generation rates for E2 and E4 as obtained from neutron, pion and proton irradiation (see text): g(E2)/g(E4) = 0.6 (5 MeV neutrons), 1.0 (192 MeV pions and 23 GeV protons) and 1.3 (27 MeV protons).

Although this so called proton-neutron puzzle is still not completely understood, the following remarks may be useful. Figure 17 shows the experimental results of point defect generation rates after n-, p- and pion irradiation [25]. The double vacancy defect E4 is surely related to cluster formation while E2 is composed of the A-center (VO$_i$) and a carbon complex (C$_i$C$_s$), both being discrete point defects. Therefore the ratio of g(E2)/g(E4) measures the point defect generation with respect to cluster formation. These findings may be compared with the simulation results shown in figure 18 [6].

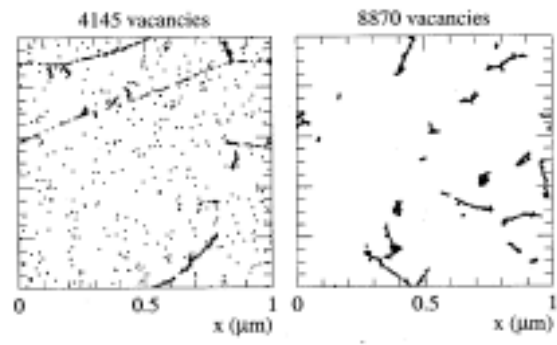

Fig. 18. Initial distribution of vacancies produced by 24GeV/c protons (left) and 1 MeV neutrons (right), corresponding to a fluence of 1·10$^{14}$cm$^{-2}$. Simulation result from Huhtinen [6].

1 MeV neutron damage is obviously dominated by the production of cluster like vacancy agglomerates with typical dimensions of 100Å or less while high energy protons lead to an additional concentration of isolated and homogeneously distributed vacancies and interstitials, responsible for point defects. The p-n-puzzle is just a result of the different primary interactions, which are in the case of neutrons caused by hard core nuclear scattering with energy transfers likely larger than the cluster formation threshold of 5 keV. For charged hadrons the Coulomb interaction prevails leading to much lower recoil energies below the cluster threshold thus responsible for more point defects. This clear picture may have to be changed if high energy pion or proton damage would be compared to that produced by neutrons with much larger energies than those accessible in reactors. In fact the neutron energy spectrum in the inner detector of LHC experiments covers a range up to about 100 MeV (figure 3). At this energy secondary nuclear reaction products as e.g. α-particles, protons and even those with larger mass could contribute to low energy recoils due to their Coulomb interaction much more than resulting from 1 MeV neutrons (elastic nuclear scattering only). Thus the p-n-puzzle could indeed be less pronounced if damage studies would be performed with higher energetic neutrons. As a result, if the DOFZ related benefits are closely related to point defect formation as it presently looks (see next section), the actual neutron damage for the tracking detectors might also be reduced by using DOFZ silicon.

### 4.5.2. microscopic picture of gamma induced damage

Although a lot of microscopic investigations on defect generation induced by hadron irradiation had been performed, it has yet not been possible to relate the macroscopic change in $N_{eff}$ to any such defects in a quantitative way. The long suspected candidate which could explain also the improvement obtained by oxygen enrichment is the acceptor like $V_2O$-center, a deep defect in the middle of the forbidden gap [41]. In contrast to hadron irradiation low energy gamma damage leads only to the generation of point defects revealing a much cleaner picture not complicated by cluster effects. Co-60 γ-irradiation studies have shown that the improvement obtained with DOFZ silicon is in this case much more pronounced than observed after hadron irradiation [42]. Only recently a defect had been measured by DLTS and TSC techniques very likely to be identified as the long searched $V_2O$ center [43, 44]. Its generation is largely suppressed in DOFZ silicon, see figure 19. Effective doping concentration does not show any type inversion as seen in standard material, see figure 20. Using the analyzed defect parameters as function of the gamma dose it was then shown that both the change in the effective doping concentration and reverse current of standard silicon detectors can largely be attributed to the generation of the $V_2O$ center [45]. This result can rightfully be regarded as an absolute first in the long and sometimes frustrating search for an understanding of macroscopic damage effects by defect analysis.

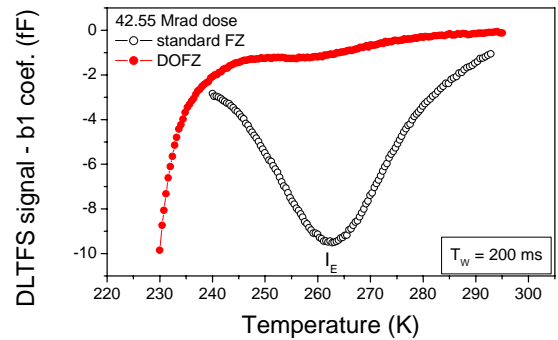

Fig. 19. DLTS spectra showing a close to midgap level at $E_a$=0.54 eV ($V_2O$) in standard silicon and suppressed in DOFZ [42, 43].

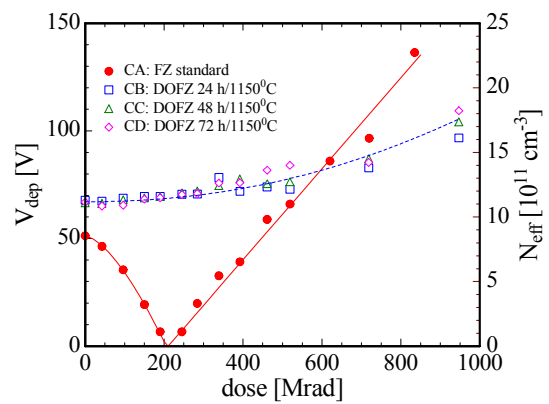

Fig. 20. Effective doping concentration as function of gamma dose for standard (closed) and DOFZ silicon (open symbols) [43].

### 4.6. Open questions

Although it looks that the final results have led to an almost universal recipe for improving the radiation tolerance in a predictable way and a possibility to perform projections to any operational scenario, a number of problems remain still open.

One of the most asked questions is: how much oxygen is enough? New results suggest that the main positive effect is already achievable with a diffusion for 24 hours at 1150°C ([20], see figures 12 and 13). However such a short diffusion process leads to a pronounced non-homogeneity in the depth profile of the O-concentration (see figure 5), the effect of which is not yet known. A detailed comparison with the BNL HTLT (<u>H</u>igh <u>T</u>emperature <u>L</u>ong <u>T</u>ime) DOFZ process resulting in a homogeneous oxygen depth profile would be very interesting [46].

Most test damage experiments have been performed with irradiations of diodes at room temperature and unbiased and hence much different from the actual situation. Only in a few cases the real operational conditions (-10°C and under bias) had been included. Examples are reported e.g. in [47], [48]. More systematic studies are needed, especially with proton resp. pion irradiation.

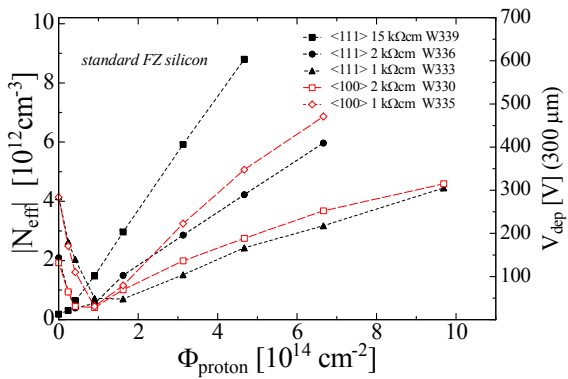

Fig. 21. Stable damage component for DOFZ detectors from different material, manufactured by ST Microelectronics (Italy).

A worrying result of principal importance is illustrated in figures 21 and 22 [49]. CERN scenario experiments with different material have shown that the "stable damage" component as function of fluence varies widely for standard FZ silicon (figure 21), whereas for DOFZ silicon the variations are

small (figure 22). The results, shown here only for diodes of one manufacturer, are much more generally

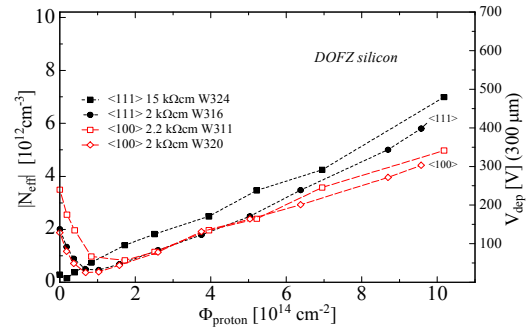

Fig. 22. Same as in Fig. 21 but for various DOFZ silicon.

Valid, including different sources. A likely reason for the effect could be a different carbon content. Reliable material characterization before and after processing will be essential.

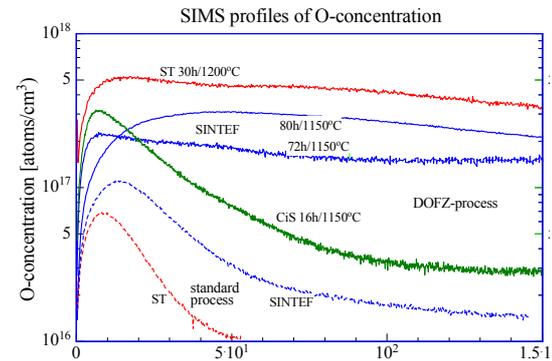

Fig. 23. Depth profiles for oxygen enrichment obtained by different DOFZ processes at various manufacturers: ST: ST Microelectronics (Italy), SINTEF (Norway), CiS (Germany). The bottom 2 curves are for a standard oxidation process, all others diffusion oxygenated as labelled.

Another riddle is displayed in figures 23 and 24 [50]. Here the correlation between the oxygen enrichment and its results for the saturation of the reverse annealing is demonstrated. In general one observes the expected effect: a larger O-concentration leads to lower reverse annealing amplitudes. E.g. the standard ST Microelectronics process (bottom most curve in figure 23) reveals the lowest O-concentration and shows the largest $N_Y$ values (top curve in figure 24). Also the ST-process with the largest O-concentration (30h at 1200°C, top curve in figure 23) results in the best reverse annealing performance (bottom most curve in figure 24). Although the CiS process (16h at 1150°C, $3^{rd}$ curve from bottom in figure 23) and the SINTEF process (72 resp. 80h at 1150°C, $4^{th}$ and $5^{th}$ from bottom in figure 23) show the expected difference in the O-concentration , the reverse annealing obtained with the CiS diodes ($2^{nd}$ from bottom in figure 24) is better than that for the SINTEF process ($3^{rd}$ and $4^{th}$ from bottom in figure 24). This quite confusing result may be understood by differences induced from the individual process. Characterization of the material after processing is therefore urgently needed, an example is given in [51].

## 5. Recent studies for improvements

The normal oxygen enrichment via the DOFZ process makes use of the diffusion of oxygen present as single atoms on interstial lattice sites. A modification had been introduced by the Brunel group generating oxygen dimers (2 closely related interstitial O-atoms). First damage studies with such material have shown promising results with evidence for the reduction of cluster related damage [52].

The BNL group, using their HTLT technique have additionally doped the material with thermal donors. An appreciable improvement had been reported showing no increase of the stable damage component of $N_{eff}$ at proton fluences above $2 \cdot 10^{14}$ p/cm² [46].

Finally the Hamburg group had performed damage tests with diodes fabricated from high resistivity Czochralski silicon, which has an as grown O-concentration of $8 \cdot 10^{17}$/cm³ (see also figure 5), much larger than possible with any DOFZ process. First damage experiments with pion and proton irradiation have shown no type inversion after large fluences, an effect which is in contrast to all FZ results and could indeed be very profitable for detector applications [20].

## 6. Conclusions

- Diffusion oxygenation of float zone silicon (DOFZ) has proven to be a powerful technique for reducing damage effects in silicon. Model parameterization of damage effects allows the calculation of detector performance for long term LHC operation.
- The DOFZ method had been successfully transferred to several detector manufacturers. It will be used for part of LHC strip and pixel detectors.
- There are still a number of open questions regarding the ROSE results. Among them are: Optimization of the O-enrichment, a possible dependence of damage effects on the initial silicon material and on the individual process technology.
- Recent updates of ROSE results have shown that the DOFZ process reveals also benefits for neutron irradiation. It is argued that these will even be larger if tests would extended to higher neutron energies, relevant for the LHC tracking area.
- An absolute first in understanding the diode performance on a microscopic level is reported for Co-60 gamma irradiated silicon.
- New studies have shown further improvements for the radiation tolerance of silicon by oxygen dimerization, dedicated doping with thermal donors and by replacing float zone with Czochralski grown silicon.
- Finally it should be mentioned that a new international collaboration had been initiated and approved by CERN as RD50, which will focus on the development of radiation hard semiconductor devices for very high luminosity colliders [53].

## Acknowledgements

The author wishes to express his thanks to the whole ROSE community for their numerous contributions, which have been reviewed in this report. I am very grateful to M. Moll (CERN) for many comments and help in collecting the presented material. The most recent results have been obtained in collaboration with CiS (Erfurt, Germany) under contract CiS-SRD 642/06/00. Finally many thanks are especially due to E. Fretwurst, I. Pintilie, J. Stahl and the whole Hamburg group for many invaluable discussions and untiring technical assistance.

# Deep defect levels in standard and oxygen enriched silicon detectors before and after $^{60}$Co-γ- irradiation[†]

J.Stahl[*], E.Fretwurst, G. Lindström, I. Pintilie[x]

*Institute for Experimental Physics, Hamburg University, Germany*

*[x]National Institute for Material Physics, Bucharest-Margurele, Romania*

**Abstract**

Capacitance Deep Level Transient Spectroscopy (C-DLTS) measurements have been performed on standard and oxygen doped silicon detectors manufactured from high resistivity n-type float zone material with <111> and <100> orientation. Three different oxygen concentrations were achieved by the so-called DOFZ process (diffusion oxygenated float zone) initiated by the CERN-RD48 (ROSE)-collaboration. Before the irradiation a material characterization has been performed. In contrast to radiation damage by neutrons or high energy charged hadrons, were the bulk damage is dominated by a mixture of clusters and point defects, the bulk damage caused by $^{60}$Co-γ-radiation is only due to the introduction of point defects. The dominant electrically active defects which have been detected after $^{60}$Co-γ-irradiation by C-DLTS are the electron traps $VO_i$, $C_iC_s$, $V_2(=/-)$, $V_2(-/0)$) and the hole trap $C_iO_i$. The main difference between standard and oxygenated silicon at low dose values can be seen in the introduction rate of $C_iC_s$ compared to $C_iO_i$. For highly oxygenated silicon the introduction of $C_iC_s$ is fully suppressed, while the sum of the introduction rates $g(C_iC_s) + g(C_iO_i)$ is independent on the oxygen concentration.

*Keywords:* radiation hardness; γ-irradiation, DLTS

## 1. Introduction

The CERN RD48 (ROSE) [1,2] collaboration had shown that oxygen enrichment of the silicon bulk leads to a considerable decrease of damage effects, induced primarily by charged hadrons and γ-irradiation. But so far it had been shown, that the observed changes of the macroscopic properties like the change of the effective doping concentration as well as the leakage current cannot be fully explained by the dominant induced defects, which have been investigated by DLTS-, TSC- and other methods [3-6].

In this work we report on DLTS-studies on $^{60}$Co-γ- irradiated silicon detectors in order to avoid the complications caused by the generation of clusters and to search for possible defects, which might be responsible for the observed changes in the macroscopic properties.

## 2 Experimental procedures

### 2.1. DLTS

The measuring method, which has been mainly used in this work, is the Capacitance-Deep-Level-Transient-Spectroscopy (C-DLTS), which was at first introduced by D.V. Lang [7]. The principle of this method is based on the temperature dependence of the emission process of trapped charge carriers in defect centers in the space charge region of a reverse biased diode. For an isolated defect level the emission process is an exponential function in time with a single emission time constant and is measured via the corresponding space charge capacitance transient. In our DLTS-system a Fast Fourier Transformation is applied to the measured transient for data reduction and evaluation of the emission time constant and trap concentration [8]. Performing a so-called temperature scan (repetition of the measurement as function of temperature) the determination of virtually all parameters associated with traps like activation energy, capture cross section, and trap concentration is possible. However the application of this method is limited by the requirement that the concentration of traps $N_T$ has to be small compared to the shallow doping concentration $N_S$. This limit can be pushed to higher trap concentrations by using the Constant-Capacitance-DLTS (CC-DLTS) technique [9]. Here the capacitance of the space charge region is kept constant by using a feedback scheme in the bias voltage-capacitance measurement system and the change of the bias voltage in the feedback loop is the DLTS-signal.

For DLTS peaks, which correspond to defect levels with similar activation energies and capture cross sections a special deconvolution technique, developed by the company PhysTech [10] had been used.

In general the CDLTS spectra had been obtained with a time window of $T_W = 200$ ms, a filling pulse duration of $t_{fill} = 100$ ms and a reverse bias of $U_R = 20$ V.

### 2.2 Material

The p+nn+ diodes investigated in this work are manufactured from high resistivity (ρ=3-4 kΩcm) n-type float zone (FZ) silicon grown in the <111> and <100> direction from Wacker. The processing and oxygen enrichment has been performed by the company CiS [11]. The oxygen doping was achieved by diffusion for 24h, 48h and 72h at 1150°C in nitrogen. This oxygenated material will be named DOFZ while the standard material will be labeled as STFZ. A detailed description of the material is given in [12].

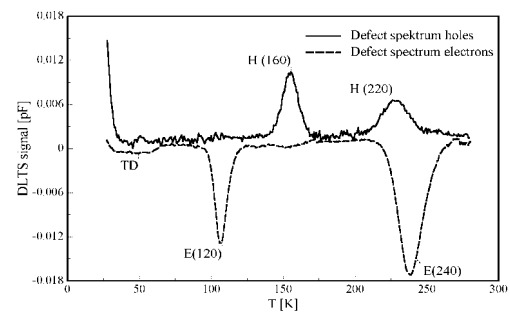

Fig. 1 CDLTS measurements of electron and hole traps in an oxygenated (24h) <100> sample before irradiation

### 2.3. Irradiation

Irradiations had been performed at the high intensity $^{60}$Co-γ-source of the Brookhaven National Laboratory (BNL) in the US at room temperature. The samples were irradiated for this investigation in a dose range from 0.2 Mrad up to 10 Mrad.

## 3. Experimental results

### 3.1. Defects before irradiation

A search for process-induced defects had been performed before irradiation. Five different defect levels have been discovered in the material.

As an example figure 1 shows DLTS spectra for electron and hole injection. The properties of these defects are listed in table 1, their concentrations are listed in table 2.

Table 1 Electric properties of the material defects

| Name | sign | $E_a$[eV] | $\sigma_{n,p}$[cm$^2$] |
|------|------|-----------|------------------------|
| TD | e | -0.137 | $1.58*10^{-13}$ |
| E(120) | e | -0.236 | $1.00*10^{-14}$ |
| E(240) | e | -0.545 | $5.41*10^{-15}$ |
| H(160) | h | +0.370 | $2.88*10^{-13}$ |
| H(220) | h | +0.494 | $1.65*10^{-14}$ |

One of these defects is the so-called thermal double donor TD. The introduction of TD's depends strongly on the oxygen concentration [13]. As can be seen in table 2 the amount of TD's is correlated with the oxygen content of the differently oxygenated devices, but the concentrations in the material with <100> orientation (CF-CH) are smaller compared to the material with <111> orientation. There is no measurable amount of thermal donors observed in STFZ-material (Wafers CA and CE).

The other four defects are so far unknown. The electron trap E(120) appears only in oxygenated <100> material. This defect has an activation energy and capture cross section which are close to the values of the double charged state of the divacancy $V_2(=/-)$ as will be shown later.

The defect E(240) is a very deep level and from its properties this level is mainly responsible for the measured generation current.

H(160) is a hole trap. It was just visible in the <100> wafer with 24h oxygenation by optical carrier injection.

H(220) is also a hole trap. It was already seen in former material investigations [5]. The concentration of this defect is independent of the oxygenation concentration and orientation of the wafer, but depends on the position of the diode on the wafer, i.e. the concentration increases with the distance from the center of the wafer to the edge. According to the material investigation the <111> material contains less defects than the <100> material.

Table 2 Concentration of material defects

| Name | CA | CB | CC | CD | CE | CF | CG | CH |
|------|-----|-----|-----|-----|-----|-----|-----|-----|
| Orientation | <111> | <111> | <111> | <111> | <100> | <100> | <100> | <100> |
| O-diff [h] | 0 | 24 | 48 | 72 | 0 | 24 | 48 | 72 |
| O [cm$^{-3}$] | | $6.2*10^{16}$ | $1.0*10^{17}$ | $1.2*10^{17}$ | | $2*10^{17}$ | $3.2*10^{17}$ | $3.1*10^{17}$ |
| C [cm$^{-3}$] | | $2.8*10^{15}$ | $5.8*10^{15}$ | $3.9*10^{15}$ | | $3.3*10^{15}$ | $3.9*10^{15}$ | $3.9*10^{15}$ |
| TD [cm$^{-3}$] | $<10^9$ | $6.0*10^9$ | $1.6*10^{10}$ | $2.4*10^{10}$ | $<10^9$ | $6.9*10^9$ | $7.2*10^9$ | $1.5*10^{10}$ |
| E (120) [cm$^{-3}$] | x | x | x | x | x | $1.7*10^{10}$ | $5.0*10^9$ | $2.0*10^9$ |
| E (240) [cm$^{-3}$] | $0.8*10^9$ | x | x | x | x | $1.2*10^{10}$ | $7.9*10^9$ | $3.7*10^9$ |
| H (160) [cm$^{-3}$] | x | x | x | x | x | $9.7*10^9$ | x | x |
| H (220) [cm$^{-3}$] | $2.4*10^9$ | $4.0*10^9$ | $5.2*10^9$ | $2.7*10^9$ | $<10^9$ | $6.3*10^9$ | $5.7*10^9$ | $1.0*10^{10}$ |

### 3.2. Radiation induced defects

The irradiation with a $^{60}$Co-$\gamma$-source has the advantage that only point defects and no clusters are generated, while an irradiation with hadrons would also cause clusters. In figure 2 a DLTS-spectrum is shown for a standard detector irradiated with a dose of 0.5 Mrad. The small peak at about 60 K is caused by the thermal double donor TD already observed before irradiation. The dominant peak at 85 K can be attributed to the creation of the $VO_i$ (vacancy + oxygen interstitial) and the $C_iC_s$(A) (carbon interstitial + carbon substitutional, state A) defects, while the small peaks at 120 K ant 200 K are associated with the different charge states of the divacancy ($V_2$(=/-) and $V_2$(-/0)). The superposition of the $VO_i$ and the $C_iC_s$ signal at 85 K cannot be resolved by high resolution techniques, since their activation energies and capture cross sections are nearly identical. But measurements of the amplitude of the DLTS signal as function of the filling pulse duration $t_{fill}$ at constant temperature allow to separate the signals arising from the $C_iC_s$(A) and $VO_i$ by making use of the bistability of the $C_iC_s$ defect[4] as demonstrated in figure 3 (upper curves). The first increase corresponds to the filling of the $VO_i$ while the second one reflects the configurational change from state $C_iC_s$(B) to state $C_iC_s$(A). For this change an activation energy has to be overcome and therefore the $t_{fill}$ value at which this step occurs depends on the temperature (see curves at T=85 K and T= 75 K). The lower curve in figure 3 was measured for an oxygenated sample where such step could not be detected. This indicates that in oxygenated material the $C_iC_s$ defect is strongly suppressed. For the extraction of the level parameters and the concentration of the double charged divacancy $V_2$(=/-) in the <100> material this defect had to be separated from the material defect that appears at the same temperature in the spectra as mentioned before. This was possible by using the high resolution DLTS method. As an example figure 4 shows the measurement of the DLTS signal as function of the transient time window (open squares) and the separation of both levels by a special refolding technique (solid line) [10]. Measurements at different temperatures result in the corresponding Arrhenius plot demonstrated in figure 5. Here level 1 corresponds to the material defect E(120) while level 2 is attributed to the $V_2(=/-)$ level. Figure 6 shows spectra for the hole trap $C_iO_i$ for samples irradiated with different doses up to 10 Mrad. As can be seen in the spectra the peak shifts slightly for higher doses to lower temperatures which is mainly caused by the fact that the concentration of this defect exceeds already at 4 Mrad the shallow doping concentration and thus the DLTS requirement is not fulfilled anymore. The activation energies and the capture cross sections of all radiation induced trap levels are summarized in table 3 and compared with values given in [4].

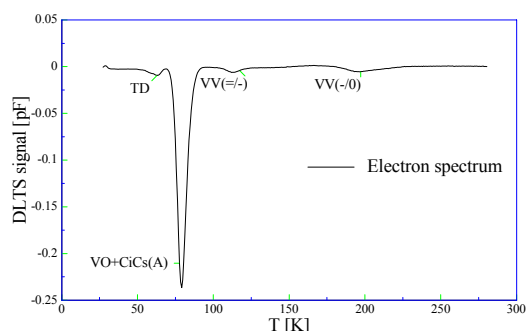

Fig. 2 CDLTS measurement of electron traps of a <100> STFZ detector irradiated with 0.5 Mrad

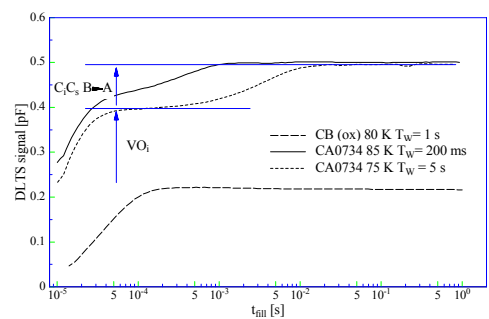

Fig. 3 Capture measurement of a standard and oxygenetaed detector irradiated with 0.5 Mrad

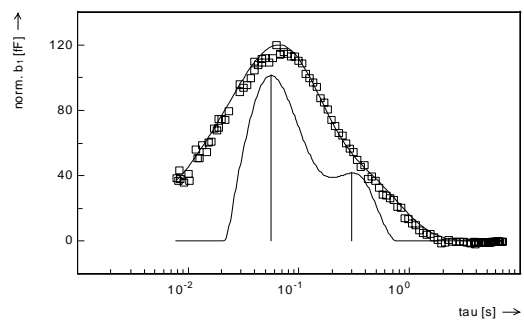

Fig. 4 Isothermal measurement at 112 K for a 24 h oxygenated <100> detector

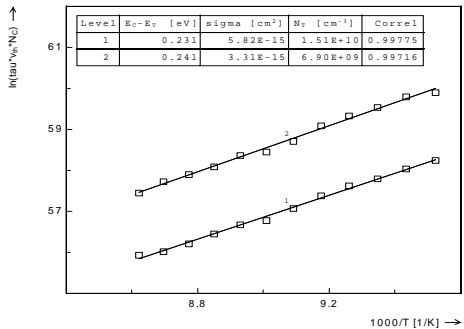

| Level | $E_t - E_v$ [eV] | sigma [cm²] | $N_t$ [cm⁻³] | Correl |
|-------|------------------|-------------|--------------|--------|
| 1 | 0.231 | 5.936e-15 | 1.518e+10 | 0.99775 |
| 2 | 0.241 | 3.318e-15 | 6.908e+09 | 0.99716 |

Fig.5 Arrhenius plot of the separation levels shown in figure 4

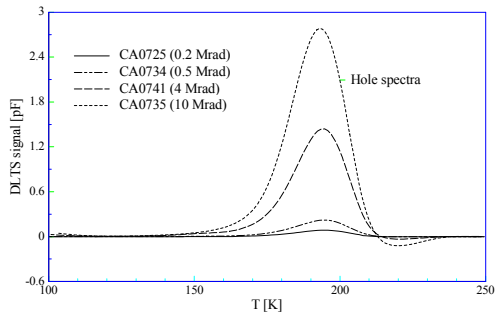

Fig. 6 CDLTS spectrum after hole injection for STFZ samples irradiated with different doses

The extracted introduction rates for the impurity related defects $VO_i$, $C_iC_s$ and $C_iO_i$ are presented in table 4 for standard and oxygenated devices (72 h) manufactured from <111> and <100> material. For the creation of $VO_i$ no dependence on the oxygen concentration is seen while the generation of $C_iC_s$ is fully suppressed in oxygenated material. This suppression is due to the fact that in oxygen rich material the probability of the reaction $C_i + O_i \rightarrow C_iO_i$ is much larger compared to the reaction $C_i + C_s \rightarrow C_iC_s$. But it is expected that the sum of the introduction rates $g(C_iC_s)+g(C_iO_i)$ stays constant for all materials. This holds with the exception observed for the standard <100> material (device CE) which shows a slightly lower introduction rate.

Table 3: Properties of radiation induced defects

| Defect | $E_a$ [eV] | $\sigma_{n,p}$ [cm²] | $E_a$ [4] [eV] | $\sigma_{n,p}$ [4] [cm²] |
|--------|-----------|----------------------|----------------|--------------------------|
| $VO_i$ | -0.172 | $1.6 \times 10^{-14}$ | -0.175 | $1.5 \times 10^{-14}$ |
| $V_2(=/-)$ | -0.251 | $3.0 \times 10^{-14}$ | -0.224 | $7.1 \times 10^{-16}$ |
| $V_2(-/0)$ | -0.426 | $3.1 \times 10^{-15}$ | -0.423 | $2.1 \times 10^{-15}$ |
| $C_iO_i$ | +0.351 | $7.3 \times 10^{-16}$ | +0.357 | $1.7 \times 10^{-15}$ |

On the other hand the introduction of divacancies $V_2$ is about two orders of magnitude smaller compared to the introduction of the impurity related defects as can be seen from table 5. A very small dependence on the oxygen concentration cannot be excluded. Furthermore, it is clear that the introduction rate for both charge states should be the same. This holds for all extracted values within the error limit of 10 %.

Table 4: Introduction rates for the impurity related defects. Values given in $10^4$rad⁻¹cm⁻³

| Name | CA stand. <111> | CD 72h <111> | CE stand. <100> | CH 72h <100> |
|------|-----------------|--------------|-----------------|--------------|
| $g(VO)$ | 70.5 | 58 | 67.3 | 80.6 |
| $g_1(C_iC_s)$ | 17.1 | 0 | 16.3 | 0 |
| $g_2(C_iO_i)$ | 47.8 | 63.4 | 33.4 | 58.5 |
| $\Sigma(g_1+g_2)$ | 64.9 | 63.4 | 49.7 | 58.5 |

It is instructive to compare the introduction rates of all $C_i$ related defects with the vacancy related ones. Assuming that the introduction of $C_i$ related defects reflects the primary generation of silicon interstitials I via the Watkins replacement mechanism I+ $C_s \rightarrow$ Si+ $C_i$ and the subsequent formation of $C_iC_s$ and $C_iO_i$ the sum of the introduction rates g($C_iC_s$) + g($C_iO_i$) should be equal to the total introduction of Interstitials g(I). On the other hand each primary interaction will create the same number of vacancies and hence the total number of vacancy related defects should balance the interstitial related ones. As a consequence the following equation for the introduction rates should be fulfilled.

Table 5: Introduction rates for the divacancies.

| Oxygenation | g ($V_2$(=/-)) $10^4$rad$^{-1}$cm$^{-3}$ | g ($V_2$(-/0)) $10^4$rad$^{-1}$cm$^{-3}$ |
|---|---|---|
| standard | 1.8 | 1.8 |
| 24 h | 1.7 | 1.5 |
| 48 h | 1.5 | 1.4 |
| 72 h | 1.5 | 1.4 |

$$\frac{g(C_iC_s) + g(C_iO_i)}{g(VO_i) + 2*g(V_2)} = 1$$

This relation holds for <111> material with a ratio of 0.91 +/- 0.12 but for <100> material the extracted ratio of 0.71 +/- 0.02 is too small. An explanation of such a small value can so far not be given.

All these defects descibed above are well known and characterized, but they still cannot explain the macroscopic properties like the increase in the leakage current and the change in the effective doping. In recent studies of our group on these devices at larger doses a close to midgap level at $E_C$-0.545 eV was observed which can mainly explain the radiation induced changes of the macroscopic detector properties[14].

## 4. Conclusions

After irradiation with $^{60}$Co-$\gamma$- radiation we observed for the different materials the dominant introduction of carbon and oxygen related defects and a much smaller generation of divacancies. The main influence of oxygen in the defect formation is seen in the formation of $C_iC_s$ and $C_iO_i$ where already oxygen enriched material by a diffusion of 24 h shows a full suppression of the $C_iC_s$ defect. In this report no difference is seen in material with different orientation. For the creation of divacancies the introduction rate is about two orders of magnitude smaller compared to the creation of impurity related defects and a very small dependence on the oxygen concentration was observed. A difference between <111> and <100> material had only been established in the ratio of introduction rates for interstitial related defects. All of these defects seen by DLTS in the investigated dose range are not responsible for the change of the macroscopic device properties. Only a recently detected deep level close to midgap can explain the macroscopic parameters.

## 5. Acknowledgments

Many thanks are due to Z. Li and E. Verbitskaja for providing the $^{60}$Co facility at Brookhaven National Laboratory and help in the gamma irradiations. The detectors were provided by CiS and the work was financed by the DFG under the contract FR 1547/1-1.

# Bulk damage effects in standard and oxygen enriched silicon detectors induced by $^{60}$Co-gamma radiation[†]

E. Fretwurst[a][*], G. Lindström[a], J. Stahl[a], I. Pintilie[a,b]
Z. Li[c], J. Kierstead[c], E. Verbitskaya[d], R. Röder[e]

[a]*Institute for Experimental Physics, University of Hamburg, 22761 Hamburg, Germany*

[b]*National Institute of Material Physics, Bucharest-Magurele, P.O. Box MG-7, Romania*

[c]*Brookhaven National Laboratory, Upton, NY 11973, USA*

[d]*Ioffe Physico-Technical Institute of Russian Academy of Sciences, St. Petersburg, Russia*

[e]*CiS Institute for Microsensors, 99099 Erfurt, Germany*

## Abstract

The influence of oxygen in silicon on bulk damage effects induced by $^{60}$Co-gamma irradiation has been studied in a dose range between 0.2 Mrad and 900 Mrad. The detector processing and oxygen enrichment were carried out in a common project by the Institute of Micro-sensors CiS using n-type high resistivity FZ silicon (3-6 k$\Omega$cm) with <111> and <100> orientation. Different oxygen concentrations were achieved by diffusion at 1150 °C for 24, 48 and 72 hours. This report on bulk damage effects is focused on the observed changes in the reverse current, the effective space charge density $N_{eff}$ extracted from C/V measurements and investigations using the transient current technique (TCT). A substantial improvement of radiation hardness concerning the development of the macroscopic properties was found for detectors manufactured on oxygenated material compared to standard material. It will be demonstrated that the change of the effective space charge density as well as the increase of the reverse current can be attributed to the creation of two deep acceptor levels and a shallow donor level.

*Keywords*: silicon detectors; oxygen in silicon; radiation damage; $^{60}$Co-gamma radiation; defects

## 1. Introduction

It was demonstrated by the CERN RD48 (ROSE) collaboration that a considerable improvement in the change of the effective space charge concentration can be achieved in oxygen enriched silicon for charged hadron damage and gamma irradiation while after exposure to neutrons no or only little suppression was found [1-3]. The beneficial effect of oxygen had been discussed and modeled under the assumption that most probably the $V_2O$ defect is responsible for a main part of the radiation induced negative space charge and that the formation of $V_2O$ is suppressed in oxygen rich material [4, 5]. The model predictions support also the experimental observation that the oxygen effect is correlated with the radiation induced formation of point defects. This means that in neutron damage the oxygen effect is suppressed since the damage is dominated by the creation of defect clusters. On the other hand a maximal oxygen effect is expected when only point defects are created which is the case in $^{60}$Co-gamma irradiation [2, 6-7]. Here we report on new results of bulk damage effects in standard and different oxygen enriched high resistivity float zone silicon with different orientation (<111> and <100>) for a wide dose range up to 900 Mrad. The experimental results for the radiation induced change of the effective doping concentration and the reverse current will be discussed in the frame work of recent microscopic studies [8,9].

## 2. Experimental procedure

The oxygen enrichment and the detector processing was carried out in a common project by the Institute for Microsensors CiS (Erfurt, Germany) using n-type high resistivity float zone (FZ) silicon from Wacker with <111> and <100> orientation. Different oxygen concentrations were achieved by diffusion at 1150 $^0$C in $N_2$ atmosphere for 24, 48 and 72 hours. The oxygen depth profiles were measured by a special SIMS-technique and compared with IR absorption measurements [10]. In Table 1 the averaged concentrations of oxygen and carbon are listed for all samples under investigation. The p$^+$nn$^+$-detectors have an active area of 0.25 cm$^2$ surrounded by a guard ring with a gap of 10μm between the outer edge of the central diode and the inner rim of the guard ring. The thickness of the different samples varies between 291μm and 295μm.

The irradiation had been performed at the high intensity $^{60}$Co-gamma source at the Brookhaven National Laboratory (BNL/USA) with a dose rate of about 600 krad per hour. The temperature during exposure was typically 27 °C. A set of 8 different samples was irradiated in consecutive steps of about 50 Mrad up to 500 Mrad and followed by larger steps up to 900 Mrad. Between each step standard I/V and C/V measurements were undertaken for the evaluation of the bulk generation current and the change of the depletion voltage $V_{dep}$ needed to fully extend the space charge region to the depth d of the diode. This voltage is related with the effective space charge density $q_0N_{eff}$ by:

$$N_{eff} = \frac{2\varepsilon\varepsilon_0}{q_0d^2}\left(V_{dep} - V_{bi}\right) \qquad (1)$$

where $\varepsilon\varepsilon_0$ is the permittivity of silicon, $q_0$ the elementary charge and $V_{bi}$ the build in voltage. Both I/V and C/V characteristics were measured at room temperature with the guard ring of the device properly connected to ground. All presented reverse current data are normalized to 20 °C.

In order to prove the reliability of the extracted values for the effective space charge density current pulse shape measurements have been performed using the transient current technique (TCT) [11,12]. For these measurements 6 sets of 8 samples each were exposed to fixed dose values.

## 3. Experimental results

Fig. 1 demonstrates the development of the depletion voltage $V_{dep}$ or effective space charge concentration $N_{eff}$ and the reverse current at total depletion $I_{rev}$ with the accumulated dose, for standard float-zone (STFZ) and three differently oxygenated float-zone (DOFZ) devices manufactured from <111> and <100> material. The main differences in the behavior of standard and oxygen enriched devices are:

Table 1

Impurity concentrations of standard and oxygen enriched devices under investigation.

| Device acronym. | CA | CB | CC | CD | CE | CF | CG | CH |
|---|---|---|---|---|---|---|---|---|
| Orientation | <111> | <111> | <111> | <111> | <100> | <100> | <100> | <100> |
| O-diffusion [h] | 0 | 24 | 48 | 72 | 0 | 24 | 48 | 72 |
| $N_{eff}$ [$10^{11}$ cm$^{-3}$] | 7.97 | 10.5 | 10.1 | 10.2 | 7.75 | 8.06 | 7.90 | 7.79 |
| [O] [$10^{16}$ cm$^{-3}$] | ≤3 | 6.2 | 10 | 12 | ≤3 | 2.0 | 3.2 | 3.1 |
| [C] [$10^{15}$ cm$^{-3}$] | <3 | 2.8 | 5.8 | 3.9 | <3 | 3.3 | 3.9 | 3.9 |

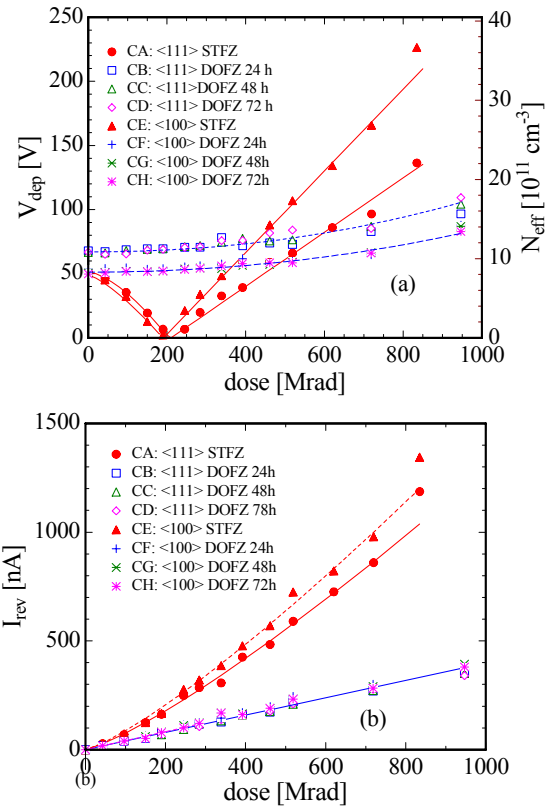

Fig 1: Dose dependence of the full depletion voltage (a) and the reverse current (b) for standard (<111>:full circle, <100> full triangle) and oxygenated (<111>:open symbols, <100>: crosses) devices.

- The detectors manufactured from standard material (CA,CE) show the well known effect of space charge sign inversion (SCSI). Initially the depletion voltage drops to a minimum value at a specific dose of about 200 Mrad before it increases again with increasing dose. This behavior is usually interpreted as a result of the creation of deep acceptor like defects and donor removal (e.g. formation of E-centers (VP$_s$)) which leads to a decrease of the initially positive space charge followed by a total compensation at a certain dose and thereafter to an inversion of the space charge sign. In contrast to this behavior, for all oxygenated devices (CB-CD, CF-CH) a small, monotonically and non-linear increase in the effective space charge density is observed. This implies that the effective space charge density becomes more positive possibly caused by an introduction of donor like defects.

- While the leakage current increase for standard material is found to be non-linear, all oxygenated detectors show a linear dose dependence. For an accumulated dose of 700 Mrad the estimated increase is about 3 times smaller compared to that of the standard devices. But no differences in the reverse current increase could be observed between the differently oxygenated sensors themselves.

The findings for the effective space charge density and in particular the speculation about the inversion had been proven by investigations of current pulse shapes recorded for injection of short (≈1 ns) 670 nm laser light pulses to the p$^+$- and the n$^+$-contact of the diodes. This way time-resolved electron and hole transport can be investigated separately. As an example Fig. 2a presents current pulse shapes of irradiated standard <111> devices for p$^+$-illumination and an applied bias voltage of 60 V that was set to be well above the full depletion voltages for this set of devices irradiated within the dose range of 43 to 284 Mrad. Since the 670 nm light is absorbed within a few microns near to the surface the presented pulse shapes are dominated by the transport of electrons in the electric field zone of the detector. In this case the negative slope of the pulse shape for doses up to 190 Mrad reflects an electric field distribution of a positive space charge and the flattening of the slope with increasing dose indicates a decrease of the space charge density as already demonstrated in the development of $N_{eff}$ extracted from C/V measurements. For dose values larger than 200 Mrad the slope switches to positive values indicating the inversion to a negative space charge. For the same dose values and illumination of the $p^+$-electrode Fig. 2b shows current pulses recorded for an oxygenated <100> sensor for a constant bias voltage of 80 V.

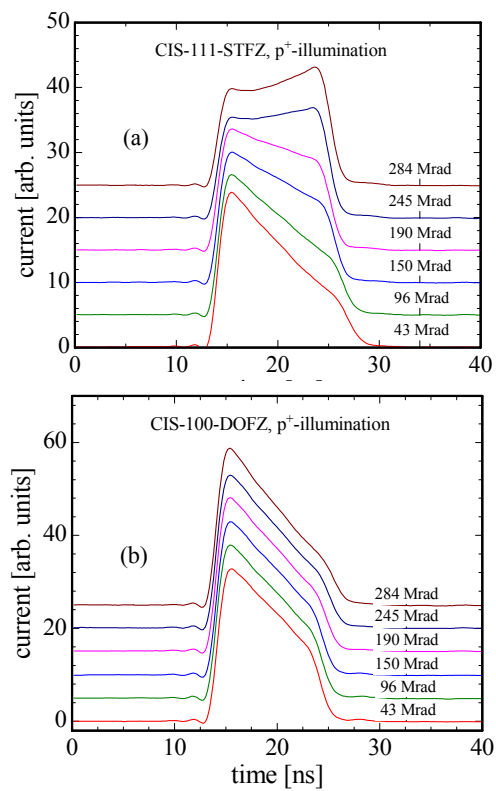

Fig. 2: Current pulse shapes after irradiation with different dose values (bottom: 43 Mrad, top: 284 Mrad) for standard <111> devices (a) and oxygen enriched <100> devices (b)

For all dose values the changes in slope of the pulses in the drift region are very small. The overall change of the slope in the range between 43 Mrad and 284 Mrad corresponds to a variation of the effective space charge concentration of $\Delta N_{eff} = 1.4 \times 10^{11}$ cm$^{-3}$ only.

## 4. Discussion

According to C. B. MacEvoy et al [7] it is assumed that the $V_2O$ defect, which was first identified in electron paramagnetic resonance (EPR) studies in heavily electron irradiated silicon [13, 14], plays the dominant role with respect to both the change of $N_{eff}$ and the generation current increase induced by $^{60}Co$ gamma radiation. In his paper it is also demonstrated that the model predicts a non-linear increase of $V_2O$ with dose. If this model is correct we should find a correlation between the reverse current increase and the change of the effective space charge concentration. In Fig. 3 the reverse current of a standard <111> detector (CA) is plotted as function of the corresponding change of the extracted effective space charge concentration defined by $\Delta N_{eff} = N_{eff,0} - N_{eff}(D)$. As indicated by the solid line we found an excellent correlation in the range below $\Delta N_{eff} = 1.3 \times 10^{12}$ cm$^{-3}$ which corresponds with the dose range up to 300 Mrad. At higher doses the current increases more rapidly than the space charge concentration which cannot be explained by an introduction of only one deep acceptor like $V_2O$. Therefore, we suspect a possible compensation of the negative space charge due to the introduction of shallow donors that are positively charged in the depletion region.

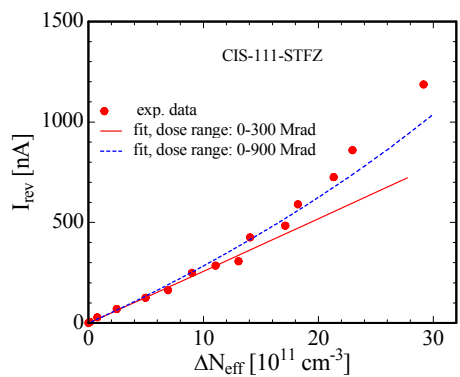

Fig. 3: Reverse current versus the change of the effective doping concentration for a standard <111> diode.

We will discuss now the experimental findings presented in chapter 3 according to recent results of detailed microscopic studies on the same material obtained by C-DLTS- and TSC-measurements [8,9]. Two close to midgap acceptor levels were discovered which are labeled as DA-I and DA-Γ. The DA-I level is most likely the $V_2O$ defect while the DA-Γ level cannot be attributed so far to a known defect. As expected for the $V_2O$ the DA-I defect is strongly reduced in oxygen rich diodes and the same behavior is also observed for the DA-Γ defect. Furthermore, in oxygen doped sensors an introduction of shallow donors (SD) had been detected in TSC-spectra. This defect is most probably the known bistable thermal double donor $TDD_2$ [15, 16]. For the calculation of $N_{eff}$ and the reverse current $I_{rev}$ as a function of dose the defect parameters and introduction rates given in [8,9] have been used.

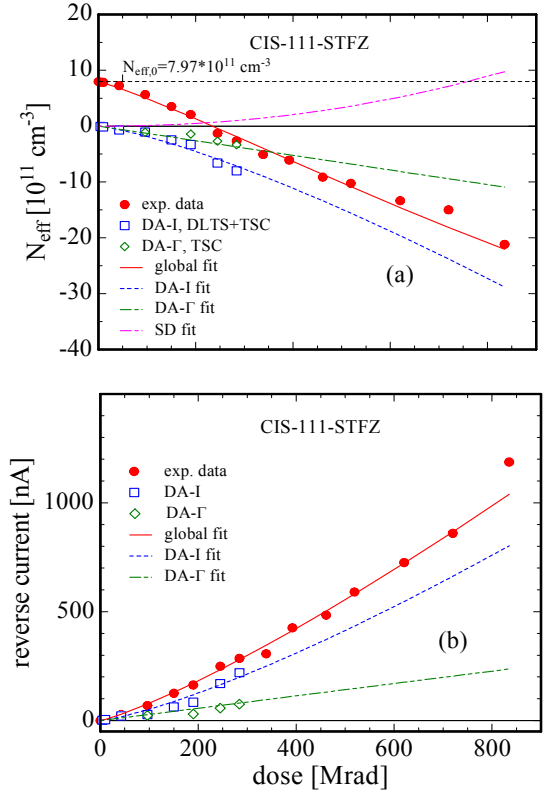

Fig. 4: Contributions of deep acceptors DA-I and DA-Γ and shallow donor SD to the effective space charge $N_{eff}$ (a) and the reverse current (b) for standard material as function of dose.

Since most of the microscopic studies were performed on irradiated samples from <111> standard and oxygenated material, we present here only calculations for the corresponding detectors taken from the same wafer.

### 4.1. Standard material

In Fig. 4a the effective space charge concentration defined by $N_{eff} = N_D - N_A$ is plotted versus the accumulated dose. Here $N_D$ represents the concentration of all donors which are positively charged in the depleted zone, i.e. the initial concentration of phosphorus and radiation induced shallow donors, and $N_A$ is the concentration of charged shallow and deep acceptors. It is obvious that both deep acceptors DA-I and DA-Γ are responsible for the inversion of the space charge sign. Their contributions to $N_{eff}$ as evaluated from DLTS- and TSC-measurements are also included in Fig. 4a. As can be seen, the dose dependence of the DA-I contribution is non-linear while that of the DA-Γ defect is linear. Taking only both acceptors into account the experimental data should follow a non-linear development in the total dose range. This is obviously not observed, moreover in the high dose range the experimental data points favor a linear dependence. Such development can only be achieved if one includes a small contribution of radiation induced shallow donors (SD), and a non-linear increase is needed in order to reproduce the experimental data. It should be mentioned that from TSC- measurements on standard material one can only find an indication for the creation of shallow donors but quantitative estimations of their concentrations were not possible [9].

Fig. 4b shows once more the experimental data for the reverse current of the standard <111> detector where the contributions of both deep acceptors are included. Also in this case the different response of the DA-I and the DA-Γ defect on the dose is clearly seen and the agreement between the measured reverse current values and the evaluated values for both defects can be stated to be excellent.

Taking these results into account, the following parameterization for the dependence of $N_{eff}$ as function of dose D was used:

$$N_{eff}(D) = N_{eff;0} + N_{SD}(D) - N_A(D) \qquad (2)$$

with $N_{eff;0} = N_{D,0} - N_{A,0}$ being the effective doping concentration before irradiation.

For the calculation of $N_{SD}$ we assume that the shallow donors are thermal double donors and, therefore, their contribution to the space charge has to be counted twice since they are double positively charged when not occupied. Furthermore, the introduction is supposed to be non-linear. This leads to the relation:

$$N_{SD}(D) = g_{SD} \times D^{\gamma_{SD}}, \qquad (3)$$

and the contribution of both deep acceptors is supposed to be:

$$N_A(D) = g_{DA\text{-}I} \times D^{\gamma_{DA\text{-}I}} + g_{DA\text{-}\Gamma} \times D \qquad (4)$$

The $g$ factors denote the effective introduction rates for the defects SD, DA-I and DA-Γ which are charged in the depleted volume (see also [9]) and the $\gamma$ values describe the non-linearity of the defect introduction as function of dose. A possible donor removal has been neglected here since the phosphorus concentration is very small and from defect kinetic simulations a very small removal constant is expected [17].

The bulk generation current $I_{rev}$ is caused by both deep acceptors only and is parameterized by:

$$I_{rev}/V = \alpha_{DA\text{-}I} \times D^{\gamma_{DA\text{-}I}} + \alpha_{DA\text{-}\Gamma} \times D \qquad (5)$$

Here V is the volume of the fully depleted detector and $\alpha$ describes the current related damage coefficients for both deep acceptor levels.

According to these parameterizations the data presented in Fig. 4a and b have been fitted. The dotted and broken lines represent the fits to the specific contributions of the defects to the effective space charge density $N_{eff}$ and the generation current $I_{rev}$ respectively. The full lines are the result of a fitting procedure which reproduces the experimental data for $N_{eff}$ and $I_{rev}$ at the same time. This procedure is denoted as "global fit".

### 4.2. Oxygen enriched material

The same fitting procedures were applied to the experimental data derived for the oxygenated device CD manufactured from <111> material. The results are shown in Fig. 5. In this case instead of $N_{eff}$ the change of the effective doping concentration defined by $\Delta N_{eff} = N_{eff}(D) - N_{eff,0}$ is plotted as function of the dose. For this oxygenated material the contribution of shallow donors could be estimated from TSC-measurements making use of the bi-stability of this defect [9]. It can be clearly seen that the shallow donors overcompensate the introduction of the deep acceptors DA-I and DA-Γ. On the other hand the introduction of both deep acceptors is strongly suppressed compared to standard material.

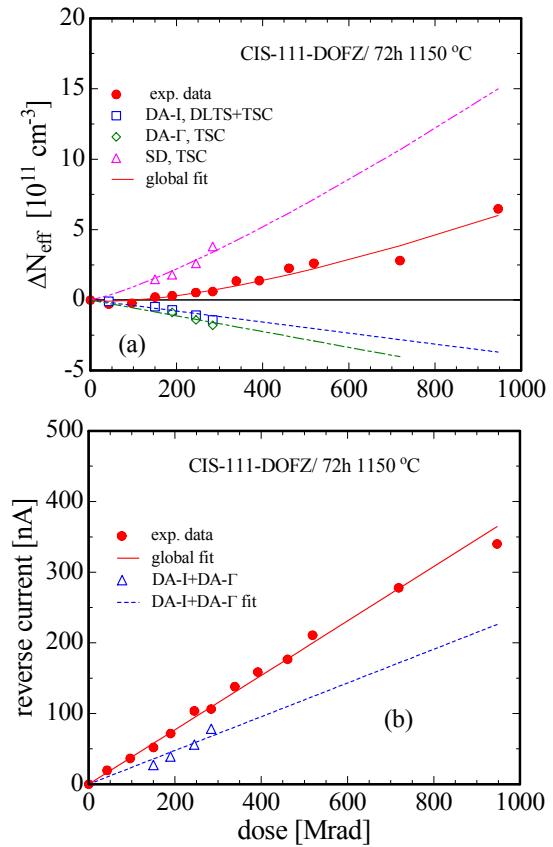

Fig. 5: Contributions of deep acceptors DA-I and DA-Γ and a shallow donor SD to the effective space charge $N_{eff}$ (a) and the reverse current (b) for oxygenated material as function of dose.

As already mentioned before the generation current increases linearly with dose (see Fig. 5b) which is expected from the observed introduction of both deep acceptors in Fig. 5a and the increase is much smaller compared to that of standard material. But the contribution of both defects to the generation current cannot fully reproduce the measured values. As pointed out in [9] possible reasons might be an underestimation of the concentration for the DA-Γ defect and/or a contribution of other deep defect centers which possibly could not be detected by TSC-measurements.

A further point of interest is the question whether any dependence on the oxygen concentration could be seen in the macroscopic parameters. From Fig. 1b we can see that the generation current is obviously not influenced by the different oxygen diffusion. This implies that already an oxygenation of 24 hours leads to the observed suppression in the creation of both deep acceptors. But an analysis of the dose dependence of $\Delta N_{eff}$ for all oxygenated detectors processed on <111> material show a very small variation of the increase with the oxygen content although the fluctuation of the data points is quite large. The result is that the introduction of shallow donors seems to be influenced by the oxygen concentration of the material but it is so far not possible to decide whether the introduction rate $g_{SD}$ or the exponent $\gamma_{SD}$ is the more sensitive parameter.

All parameters of the parameterization given by eq.(3-5) which had been evaluated from the described fitting procedures are summarized in Table 2.

Further studies of these and other differently processed detectors were perfomed for irradiations up to ultra-high dose values between 0.9 and 1.76 Grad. The results are described separately in [18].

Table 2

Extracted parameters according to eq.(3-5); g-values are given in units [$10^8$ cm$^{-3}$ Mrad$^{-1}$] and $\alpha$ values in [$10^{-8}$ Mrad$^{-1}$].

| Device | CA | CB | CC | CD |
|---|---|---|---|---|
| Oxygenation | Standard | 24 h | 48 h | 72 h |
| $g_{DA-I}$ | 4.80 | 3.90 | 3.90 | 3.90 |
| $\gamma_{DA-I}$ | 1.29 | 1.0 | 1.0 | 1.0 |
| $g_{DA-\Gamma}$ | 1.31 | 0.559 | 0.559 | 0.559 |
| $g_{SD}$ | 0.0065 | 5.45 | 5.59 | 3.25 |
| $\gamma_{SD}$ | 2.12 | 1.12 | 1.14 | 1.23 |
| $\alpha_{DA-I}$ | 1.83 | 1.48 | 1.48 | 1.48 |
| $\alpha_{DA-\Gamma}$ | 3.90 | 3.76 | 3.76 | 3.76 |

## 5. Conclusions

The presented study on bulk damage effects after $^{60}$Co gamma irradiation up to 900 Mrad had substantiated the strong influence of oxygen in silicon for the development of more radiation tolerant detectors. For the first time the observed changes of the macroscopic detector properties concerning the effective doping concentration as well as the reverse current can be explained by radiation induced deep acceptors and the creation of shallow donors which had been discovered in microscopic studies.

## Acknowledgements

This work has been performed in the frame work of the CiS-SRD project under contract 642/06/00. Financial support of the German Research Foundation DFG under contract FR1547/1-1 and partly by the German Ministry for Education and Research BMBF under contract WTZ-ROM 00/01 is gratefully acknowledged.

# Results on defects induced by $Co^{60}$-gamma irradiation in standard and oxygen enriched silicon[†]

I. Pintilie[a)b)][*] E. Fretwurst [b)], G. Lindström [b)] and J. Stahl [b)]

[a)] *National Institute of Materials Physics, Bucharest-Magurele, P.O.Box MG-7, Romania*

[b)] *Institute for Experimental Physics, Hamburg University, D-22761, Germany*

**Abstract**

Radiation induced defects in silicon diodes were investigated after exposure to high doses of Co-60 gamma irradiation, using Deep Level Transient Fourier Spectroscopy and Thermally Stimulated Current methods. The main focus was on differences between standard and oxygen enriched material and the impact of the observed defect generation on the diode properties. Two close to mid gap trapping levels and a bi-stable donor level have been characterized as function of dose. These defects explain the main macroscopic deterioration effects both in standard and oxygen enriched float zone diodes. Radiation damage effects in silicon detectors under severe hadron- and $\gamma$–irradiation are surveyed, focusing on bulk effects. Both macroscopic detector properties (reverse current, depletion voltage and charge collection) as also the underlying microscopic defect generation are covered. Basic results are taken from the work done in the CERN-RD48 (ROSE) collaboration updated by results of recent work. Preliminary studies on the use of dimerized float zone and Czochralski silicon as detector material show possible benefits. An essential progress in the understanding of the radiation induced detector deterioration had recently been achieved in gamma irradiation, directly correlating defect analysis data with the macroscopic detector performance.

*Keywords:* silicon detectors; defect emgineering; point defects defect analysis

## 1. Introduction

High resistivity silicon particle detectors will be used extensively in the tracking areas of the CERN Large Hadron Collider (LHC) experiments. Since for this application the detectors are subject to extremely high hadron fluences the radiation tolerance of the bulk material became of prime importance. In all cases high resistivity silicon ($\rho > 1$ kΩcm) was chosen to allow for low operating voltages.

A lot of investigations on hadron induced damage and the impact on detector performance had been extensively studied by the CERN RD48 (ROSE) collaboration [1-6]. It had especially been proven that the deliberate addition of oxygen ([O]>$10^{17}$ cm$^{-3}$) in the bulk material is beneficial for the radiation tolerance to charged particle and gamma exposure. The clearest benefit is gained in the case of $^{60}$Co gamma irradiation. No change in the initial doping concentration was reported up to a dose of 400 Mrad for high resistivity Diffusion Oxygenated Float Zone (DOFZ) silicon while in the case of Standard Float Zone (STFZ) silicon the type inversion takes place around 250 Mrad [2]. Many radiation induced point defects (VV$^{=/-}$, VV$^{-/0}$, C$_i$C$_s^-$, VO$_i^{-/0}$, C$_i$O$_i^{+/0}$ etc. ) were already well characterized by various methods [5-17]. However, these defects cannot explain the irradiation-induced change in the macroscopic parameters of the detector performance (doping concentration and leakage currents) and the improvement of the radiation tolerance for oxygen-enriched silicon. Present defect models attribute the beneficial oxygen effect in gamma irradiated DOFZ silicon to a lower probability of V$_2$O formation by an enhanced production of VO defects [17-20]. The V$_2$O defect was identified in Czochralski (CZ) silicon after electron irradiation by Electron Paramagnetic Resonance (EPR) [21] and Photo-EPR measurements [20]. An activation enthalpy of (0.50±0.05) eV from the conduction band was evaluated for the V$_2$O defect in the neutral charge state [22]. Recently, we have reported on a deep level located at 0.545 eV from the conduction band (E$_c$) and having electron and hole capture cross sections of $\sigma_n = (1.7\pm0.2)\times10^{-15}$ cm$^2$ and $\sigma_p = (9\pm1)\times10^{-14}$ cm$^2$ respectively detected in float-zone silicon after Co$^{60}$-gamma irradiation with doses up to 42.5 Mrad [23]. This close to midgap trap (the I level in ref.[23]) is strongly generated in STFZ diodes but strongly suppressed in DOFZ devices and was suggested to be related to the V$_2$O complex in the single charge state. According to our results it can explain about 90% of the change in the effective doping concentration (Neff) and 50% of the leakage current (LC) in STFZ silicon. Considering that the I level was also detected in DOFZ material together with the still unexplained difference in the measured and calculated leakage current call for further investigations, especially after higher doses of irradiation. In the following we report on the detection by Deep Level Transient Fourier Spectroscopy (DLTFS) [24] and Thermally Stimulated Current (TSC) [13] methods of other trapping levels induced by gamma irradiation, which together with the mentioned I level can better explain the differences observed between STFZ and DOFZ diodes. The devices investigated in this work are STFZ and DOFZ p$^+$nn$^+$ diodes processed by CiS [25] on Wacker silicon with high resistivity (4 kΩcm) and exposed to Co$^{60}$- gamma irradiation doses of 10 to 300 Mrad. The average oxygen concentration in the DOFZ material is 1.2x$10^{17}$ cm$^{-3}$ while in the STFZ ones it is less than 3x$10^{16}$ cm$^{-3}$ [26].

## 2. Experimental results and discussions

### 2.1. DLTFS results

The DLTFS method was applied for doses up to 42.5 Mrad whereas the higher irradiated samples were investigated by the TSC method. After irradiation all dides were kept at room temperature for at least 4 weeks. DLTFS measurements were performed on both STFZ and DOFZ samples exposed to 10 and 42.5 Mrad irradiation doses. For these high doses the DLTFS spectra can be well analyzed only for temperatures above 240 K where the most abundant irradiation induced point defects (e.g. VO and VV) do not contribute anymore to the recorded capacitive transients. In addition to the mentioned I level (possibly the V$_2$O$^{-/0}$ complex) another very deep level, labeled as Γ, was detected in these samples as shown in Fig. 1a. An activation enthalpy of ΔH$_v$ =

0.66eV ±1% with respect to the valence band and an effective capture cross-section of σ* = (5±3)x10⁻¹⁵ cm² were evaluated for the Γ defect from the Arrhenius plot (Fig. 1b). The filling was done using forward bias injection and therefore σ* does not represent the capture cross section for holes only ($\sigma_p$) but includes also the electron capture cross section ($\sigma_n$). The σ* is given by the following formula [27]:

$$\sigma^* = \sigma_p + \sigma_n \frac{v_{th,n}}{v_{th,p}} \qquad (1)$$

where $v_{th,p}$, $v_{th,n}$ are the thermal velocities of holes and electrons respectively.

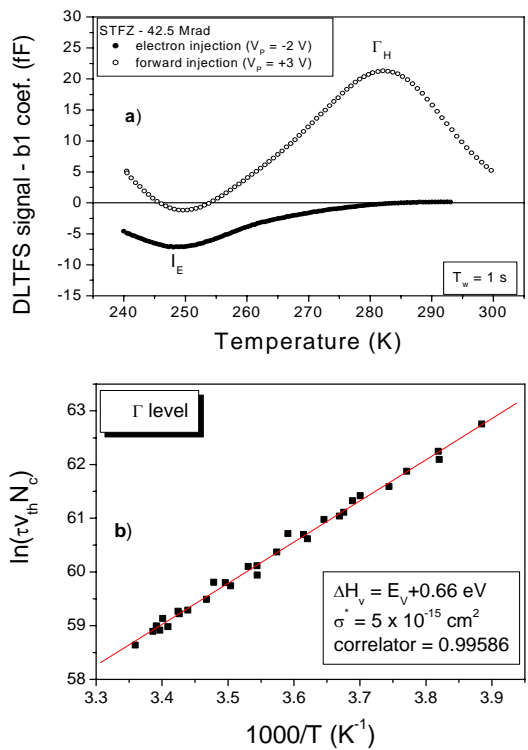

Fig. 1 a) Majority and minority carrier DLTFS spectra recorded on STFZ diode after a gamma dose of 42.5 Mrad. A reverse bias of RB=-16 V was applied during the measurement. b) Arrhenius plot for the Γ level with the evaluation of the activation enthalpy and capture cross section.

The concentration evaluated from DLTFS spectra or by direct analysis of transients after forward injection does not represent the total concentration of defects ($N_T$) but a quantity given by [27]:

$$p_T^{DLTS}(T) = (N_T - n_T(T)) * \frac{c_p(T)}{c_p(T) + c_n(T)}$$

with

$$n_T(T) = N_T * \frac{c_n(T) * n + e_p}{e_n(T) + e_p(T) + c_n(T) * n + c_p(T) * p} \qquad (2)$$

$$e_{n,p}(T) = c_{n,p}(T) * N_{C,V}(T) * \exp(\pm \frac{E_a(T) - E_{C,V}}{k_b T})$$

where $n_T(T)$ is the steady state occupancy of the Γ level, $c_{n,p}(T) = \sigma_{n,p}(T) * v_{th,n,p}(T)$ are the respective capture coefficients, $N_{C,V}$ are the effective densities of states in the conduction band.

Due to its position in the band gap, the Γ level might have a strong influence on the detector performances – the leakage current per unit volume (LC) and the effective doping concentration ($N_{eff}$). In order to estimate it correctly both capture cross sections ($\sigma_n$ and $\sigma_p$) should be known. Because of the high concentration of VV⁻/0 states a direct investigation of electron emission from the Γ defect is not possible. By direct analysis of transients at different temperatures the $p_T^{DLTS}(T)$ values were determined to be 9.60x10⁹cm⁻³, 9.44x10⁹cm⁻³ and 9.36x10⁹cm⁻³ at 273 K, 293 K and 298 K respectively. Assuming that there is no temperature dependence of $\sigma_p$ and $\sigma_n$ between 273 K and 298 K then eq. 2 results in $\sigma_p/\sigma_n \sim 100$. With such a large difference between the two capture cross sections a good approximation is $\sigma_p = \sigma^* = (5\pm3)x10^{-15}$ cm² as was evaluated from the Arrhenius plot in fig.1b.

## 2.2. TSC results

The samples irradiated with higher doses (up to 284 Mrad) were investigated by the TSC method. Also in this case the diodes stayed at room temperature for more than 4 weeks prior to the measurements. The standard experimental procedure consists in cooling down from room temperature (RT) to 20 K under 0 bias, followed by forward injection (1mA) for 30 sec. and then heating up with reverse bias (RB) applied. The heating rate for the TSC experiments was always 0.183 K/s. The filling of traps was done either by forward biasing or using hole injection by 670 nm LED illumination from the rear side of the reverse biased diode.

In Fig. 2 the recorded TSC spectra for STFZ and DOFZ diodes exposed to a dose of 284 Mrad are presented. In DOFZ diodes the generation of I and Γ defects are strongly suppressed, while the VOi center exhibits a larger concentration. This may suggest that the $VO_i$ center plays a major role in the formation of I and Γ defects. In fact, it had been discussed that the $V_2O$ complex results from the reaction $VO_i+V \Rightarrow V_2O$ [17-20]. In addition, there are some features characteristic only for the STFZ or DOFZ material neither reported so far in the literature nor seen in our low dose DLTFS measurements [28].

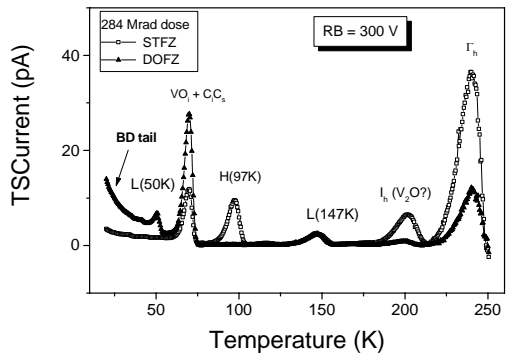

Fig. 2. TSC spectra for STFZ and DOFZ diodes irradiated with 284 Mrad dose recorded after exposure to day light.

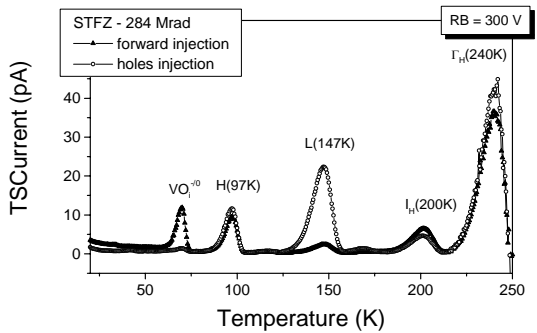

Fig. 3. TSC spectra in STFZ for only hole injection or after forward biasing.

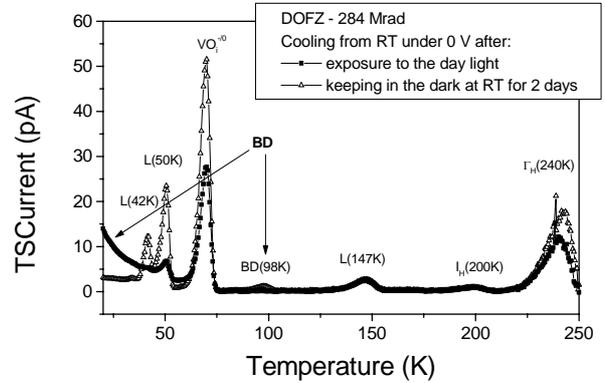

Fig. 4. TSC spectra in DOFZ diodes after forward biasing following different exposures.

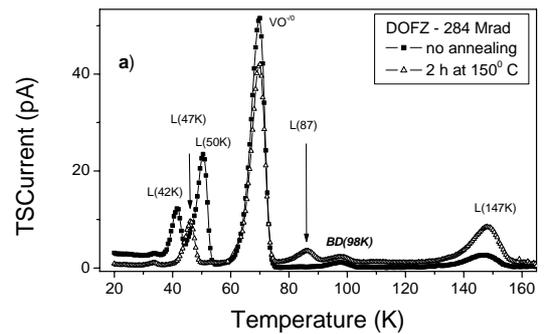

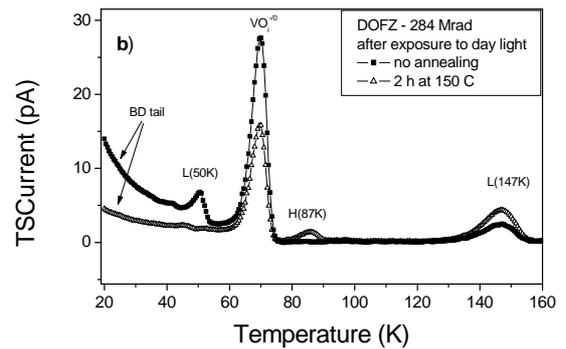

Fig. 5. Annealing at 150 $^0$C . a) TSC spectra for DOFZ after keeping the diode in the dark at RT for 2 days; b) TSC spectra for DOFZ after exposure to day light

In the STFZ material a strong peak is observed at 97 K which proved to act as a trapping level for holes (see Fig. 3). In the DOFZ case we see a peak at 50 K overlapping a very low temperature signal labeled "BD tail". The BD tail disappears after keeping the diode in the dark for 2 days at room temperature (RT). Instead, the L(42K) and BD(98K) levels appear in the TSC spectrum (see Fig. 4). A level with similar trapping parameters like L(50K) was already investigated by DLTS and IR- Absorption methods in oxygen rich Si and was associated with the single negative charge state of the interstitial silicon-oxygen dimer complex $(IO_2^{-/0})$ annealing out completely at 150 °C [15]. Our experiments revealed that both L(50K) and L(42K) levels anneal out after 150° C heat treatment (Fig. 5). Instead of these two levels another peak labeled L(87K) appears in the spectrum. The L(47K) peak is not a result of the annealing. It can already be seen as a shoulder in the L(50K) peak prior to annealing.

It is worth to mention here that no change in the magnitude of the leakage current or Neff was measured after the annealing at 150° C. Depending on the experimental TSC procedure (cooling after exposure to day light or after keeping the sample in the dark) a direct correlation between the BD tail and BD(98K) level was observed (see Fig. 5). It is well known that such bi-stability is characteristic for the thermal double donors TDD1 and TDD2 (associated with small size oxygen clusters) [29-31].

The donor activity of the BD(98K) level was proven observing the Poole Frenkel effect when different RB-values were applied during the TSC measurements. The change in the activation enthalpy was evaluated from Arrhenius plots of the increasing part of the peak (see Fig. 6) resulting in a zero field activation enthalpy of 0.225 eV for BD(98K) . Thus, contrary to the effect of I and Γ the BD centers will contribute with positive space charge to Neff. These properties of the BD centers (donor activity, bi-stability, zero field activation energy) together with their strong generation in oxygen enriched material suggest a possible identification with the thermal double donors TDD2.

## 3. The impact of I, Γ and BD defects on the detector performance

The following relations were used to calculate the contribution of the I, Γ and BD centers to LC ($\alpha_E$) and $N_{eff}$ ($n_T$):

$$\alpha_E(T) = q_0 * e_n(T) * n_T(T)$$
and
$$n_T(T) = N_T * \frac{c_n(T) * n + e_p}{e_n(T) + e_n(T) + c_n(T) * n + c_n(T) * p} \quad (3)$$

where $q_0$ is the free electron charge.

The concentrations of I, Γ and BD centers in STFZ and DOFZ for different dose values are given in Tab. I. The errors are less than 5% in the case of I level and around 20% for the Γ level.

The experimental values of $N_{eff}$ and LC as determined from C-V and I-V measurements as well as the calculated ones are given in Fig. 7. There is a very good agreement between measured and calculated values of Neff for both STFZ and DOFZ diodes. However, the LC cannot be fully described by taken only the I and Γ level into account (especially in the case of DOFZ). Possible reasons for that can be that other deep centers (e.g. $VV^{-/0}$) have not been included.

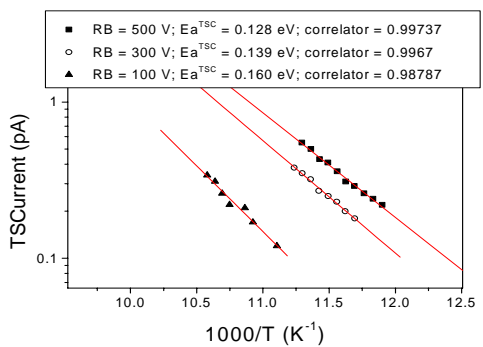

Fig. 6. Change of the activation enthalpy of BD(98K) with different reverse bias applied during TSC measurement.

Table I. Total concentration of I, Γ and BD defects as function of irradiation dose in STFZ and DOFZ silicon

| Dose (Mrad) | I level STFZ x10⁹ (cm⁻³) | I level DOFZ x10¹⁰ (cm⁻³) | Γ level STFZ x10¹² (cm⁻³) | Γ level DOFZ x10¹² (cm⁻³) | BD DOFZ x10¹¹ (cm⁻³) |
|---|---|---|---|---|---|
| 4 | 6.08 | | | | |
| 10 | 11.45 | | | | |
| 42 | 62 | 1.6 | | | |
| 96 | 120 | | 1.9 | | |
| 150 | 276 | 5 | 3.2 | | 1.48 |
| 190 | 380 | 7.5 | 2.4 | 1.5 | 2.04 |
| 245 | 700 | 11.5 | 5 | 2.1 | 2.96 |
| 284 | 920 | 16 | 6 | 3 | 3.8 |

| Dose (Mrad) | I level STFZ $\times 10^9$ (cm$^{-3}$) | I level DOFZ $\times 10^{10}$ (cm$^{-3}$) | Γ level STFZ $\times 10^{12}$ (cm$^{-3}$) | Γ level DOFZ $\times 10^{12}$ (cm$^{-3}$) | BD DOFZ $\times 10^{11}$ (cm$^{-3}$) |
|---|---|---|---|---|---|
| 4 | 6.08 | | | | |
| 10 | 11.45 | | | | |
| 42 | 62 | 1.6 | | | |
| 96 | 120 | | 1.9 | | |
| 150 | 276 | 5 | 3.2 | | 1.48 |
| 190 | 380 | 7.5 | 2.4 | 1.5 | 2.04 |
| 245 | 700 | 11.5 | 5 | 2.1 | 2.96 |
| 284 | 920 | 16 | 6 | 3 | 3.8 |

## 4. Conclusions

The DLTFS method applied to gamma irradiated STFZ and DOFZ silicon diodes with doses up to 42 Mrad allowed us to characterize a new close to midgap trapping level (Γ center) in addition to the already reported I defect ($V_2O^{-/0}$?).The trapping parameters of the Γ level are: an activation enthalpy of $\Delta H_v = 0.66$eV $\pm 1\%$ from the valence band and a capture cross-sections of $\sigma_p = 100 \times \sigma_n = (5\pm3) \times 10^{-15}$ cm². The dependence of the I and Γ concentrations with increasing dose was investigated by the TSC method. It was shown that both defects are largely suppressed in DOFZ diodes. In STFZ the Γ level can almost fully explain the change in both the leakage current and the effective doping concentration . In the case of DOFZ diodes the negative space charge due to I and Γ centers is overcompensated by the positive charge introduced by some bi-stable donors (TDD2 ?) leading even to a slight increase of the effective donor concentration.

### Acknowledgements

Many thanks are due to Z. Li and E. Verbitskaja for help in the gamma irradiations at Brookhaven National Laboratory. This work has been performed in the frame of the CiS-SRD project under contract 642/06/00. Financial support of the German Ministry for Education and Research BMBF and of the Romanian Ministry for Education and Research under contract WTZ-ROM 00/01 and of the German Research Foundation DFG under contract FR1547/1-1 is gratefully acknowledged.

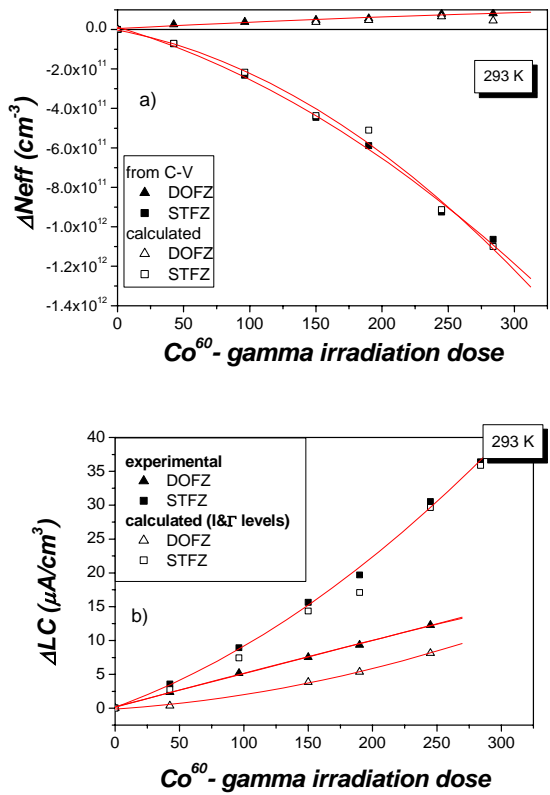

Fig. 7. Experimental and calculated dose dependence of: a) effective doping concentration
b) leakage current per unit volume